\documentstyle[apj,times,psfig]{article}

\def\simg{\mathrel{\rlap{\raise 0.511ex \hbox{$>$}}{\lower 0.511ex \hbox{$\sim$}}}}
\def\siml{\mathrel{\rlap{\raise 0.511ex \hbox{$<$}}{\lower 0.511ex \hbox{$\sim$}}}}
\def\veps{\varepsilon} \def\eps{\epsilon} 
\def\tveps{\tilde{\varepsilon}} \def\teps{\tilde{\epsilon}} \def\tGamma{\tilde{\Gamma}}
\def\ergcm2{\, \rm erg \, cm^{-2}} \def\fluxcgs{\, \rm erg \, cm^{-2} s^{-1}} 
\def\taugg{\tau_{\gamma\gamma}} \def\fgg{f_{\gamma\gamma}} \def\Z{{\cal Z}}
\def\N{{\cal N}}

\begin{document}

\parskip 5pt

\title{Optical Flashes from Internal Pairs Formed in Gamma-Ray Burst Afterglows}

\author{A. Panaitescu}
\affil{ Space \& Remote Sensing, MS B244, Los Alamos National Laboratory, Los Alamos, NM 87545, USA}

\begin{abstract}
 We develop a numerical formalism for calculating the distribution with energy of the (internal) pairs 
formed in a relativistic source from unscattered MeV--TeV photons. 
 For GRB afterglows, this formalism is more suitable if the relativistic reverse-shock that energizes 
the ejecta is the source of the GeV photons. 
 The number of pairs formed is set by the source GeV output (calculated from the Fermi-LAT fluence), 
the unknown source Lorentz factor, and the unmeasured peak energy of the LAT spectral component. 
 We show synchrotron and inverse-Compton light-curves expected from pairs formed in the shocked medium 
and identify some criteria for testing a pair origin of GRB optical counterparts. 
 Pairs formed in bright LAT afterglows with a Lorentz factor in the few hundreds may produce bright 
optical counterparts ($R < 10$) lasting for up to one hundred seconds. 
 The number of internal pairs formed from unscattered seed photons decreases very strongly with the 
source Lorentz factor, thus bright GRB optical counterparts cannot arise from internal pairs if
the afterglow Lorentz factor is above several hundreds. 
\end{abstract}


\vspace*{2mm}
\section{Introduction}
 
 The LAT instrument onboard the Fermi satellite has detected a high-energy emission at 100 MeV--100 GeV, 
extending well after the GBM prompt phase, for dozens of GRB afterglows (first Fermi catalog --
Ackermann et al 2013). The properties of the LAT afterglow emission are: fluence above 100 MeV 
$\Phi = 10^{-5\pm 1} \ergcm2$, light-curve peak at 10--20 s after trigger, post-peak flux decay 
$\nu F_\nu \propto t^{-1.3\pm 0.3}$ monitored up to 1 ks (sometimes longer), photon spectrum $C_\nu 
\propto \nu^{-2.1\pm 0.2}$. 

  The isotropic energetic output of the brightest LAT afterglows, $E_\gamma = 10^{53 \pm 1}$ erg, 
is 10--100 percent of the GRB output at $\siml 1$ MeV, the corresponding number of afterglow photons 
above $\varepsilon_\gamma \sim 100$ MeV being $N_\gamma = E_\gamma/\varepsilon_\gamma \simeq 
10^{57 \pm 1}$. The fraction $f_\pm$ of these photons that form pairs depends strongly on 
the Lorentz factor $\Gamma$ of the medium that produced the LAT afterglow emission, because 
$\Gamma$ determines the lab-frame collimation of photons and the threshold energy for pair-formation, 
and on the source radius $R$, which sets the optical-thickness to photon-photon absorption. 
Taking into account that $R \simeq \Gamma^2 ct$, with $t$ the observer time, it follows that 
$f_\pm$ has a very strong dependence on $\Gamma$.
 
 From the escape of the higher-energy LAT photons ($\sim 10$ GeV), Abdo et al (2009) have set lower
limits $\Gamma_o \simg 200-1000$ for the Lorentz factor of several LAT sources during the prompt emission
phase (burst). Consistent with that, Panaitescu et al (2014) found that, for $\Gamma \simg 200$, 
photon-photon attenuation does not yield a spectral signature but, for $\Gamma \siml 200$, 
the attenuation of photons above 1 GeV should be detectable. Additionally, as shown in this article, 
for $\Gamma \siml 500$, the number of pairs $N_\pm = f_\pm N_\gamma$ is higher than the number 
of electrons energized by the forward shock and, for $\Gamma \siml 250$, is larger than the number of
ejecta electrons energized by the reverse shock. Therefore, the formation of pairs from LAT photons
is of importance at least for those GRBs/afterglows whose LAT spectrum displays a spectral softening 
at GeV energies.

 Assuming a single power-law for the LAT spectral component, Panaitescu \& Vestrand (2014) have 
calculated {\sl analytically} the emission from the pairs formed only from photons to which the GeV 
front is optically-thick, leading to the conclusion that pairs can account for the brightest 
optical counterparts (flashes) observed during the prompt phase. A larger number of pairs, but of lower 
energy, are formed by the photons for which the LAT emission is optically-thin (to pair-formation). 
In this work, we calculate {\sl numerically} the distribution with energy of all pairs formed from 
high-energy photons (assuming a broken power-law spectrum), integrate it over the deceleration of 
the blast-wave that produces the LAT afterglow, and track numerically the pair radiative cooling, 
to obtain accurate light-curves for the optical flashes produced by pairs formed in GRBs and afterglows.

 The effects arising from scattering of the high-energy photons on (cold) electrons existing in the
source or on the already-formed pairs are ignored. Such scattering increases the source-frame photon 
escape path, which increases the probability that any photon forms a pair and the total number
of pairs formed. As we shall see, for sources that are optically-thin to photon scattering, scattering
on the already-formed pairs occurs with a smaller probability than pair-formation thus, to a good
approximation, the effect of photon scattering on the pair-formation rate can be ignored. The same is true for
scattering on the ejecta electrons, if the Lorentz factor of the GeV source is in the few hundreds.
However, if that Lorentz factor is or exceeds several hundreds, scattering on ejecta electrons should
occur with a higher probability than pair-formation. In this case, the number of pairs formed from
unscattered seed photons, calculated below, underestimates the true number of pairs.

 We also ignore the formation of (external) pairs ahead of the afterglow blast-wave, which loads 
with leptons the ambient medium and accelerates it (Beloborodov 2002), changing the afterglow dynamics 
(Kumar \& Panaitescu 2004).
Consequently, the formalism presented here for the emission from pairs is more pertaining to a GeV
source that is located well behind the forward-shock, so that most pairs form in the shocked fluid
and not ahead of the blast-wave. That condition points to a relativistic reverse-shock as the origin
of the LAT afterglow emission (as could be the case for GRB 130427A - Panaitescu et al 2013).

\section{Pair-formation in a relativistic source}

\subsection{High-energy spectral component}

 The number of pairs formed at any observer-frame time $t$ over a dynamical timescale is derived
from the observable 0.1--10 GeV fluence $\Phi$, the spectrum of the high-energy afterglow emission
(with the 0.1-10 GeV spectral slope being the only observational constraint), the source redshift $z$,
and the unknown Lorentz factor $\Gamma$ of the high-energy source. At the redshift of the afterglow source, 
the afterglow photon spectrum is assumed to be a broken power-law:
\begin{equation}
 \frac{dN_\gamma}{d\eps} =  \left. \frac{dN_\gamma}{d\eps}\right|_{\eps_b}  \left\{ \begin{array}{ll}  
      (\eps/\eps_b)^{-\alpha} & \eps < \eps_b \\ (\eps/\eps_b)^{-\beta} & \eps_b < \eps \\
      \end{array} \right.
\label{dNg}
\end{equation}
where $\eps_b$ is the spectral-break energy at redshift $z$ (i.e. the peak of the luminosity $\nu L_\nu$
spectrum), $\alpha$ and $\beta$ being the low and high-energy spectral slopes. For synchrotron and
inverse-Compton emissions, $\alpha$ has four possible values: 2/3 for optically-thin synchrotron (sy)
or inverse-Compton (ic) emission from un-cooled electrons, 3/2 for optically-thin sy/ic from cooled 
electrons (i.e. with a radiative cooling timescale shorter than the age of the source), -1 for 
self-absorbed synchrotron emission (but is unlikely that the source magnetic field is sufficiently
high for self-absorption to be important at MeV), and 0 for the inverse-Compton scattering of 
self-absorbed synchrotron emission. 
Then, the spectral slope around 2 measured by LAT above 100 MeV indicates that $\veps_b < 100$ MeV
(in the observer frame), $\beta \simeq 2$, and that the low-energy spectrum is not observed, 
being dimmer at 10 keV--10 MeV than the GRB spectrum. 

 The normalization factor of equation (\ref{dNg}) is simply set by the measured fluence $\Phi$ 
\begin{displaymath}
 \Phi (0.1-10 {\rm GeV}) = \frac{(z+1)^3}{4 \pi d_l^2} \int_{0.1GeV}^{10GeV} 
   d\veps \, \veps \left.\frac{dN_\gamma}{d\eps}\right|_{(z+1)\veps}  
\end{displaymath}
\begin{equation}
   =  \frac{(z+1)^3}{4 \pi d_l^2} f(\beta) \veps_b^2 \left. \frac{dN_\gamma}{d\eps}\right|_{\eps_b}
\label{phi}
\end{equation}
where $d_l \simeq 5.10^{52} (z+1)^2$ cm is the luminosity distance and
\begin{equation}
   f(\beta) = \frac{1}{\beta-2} \left[ \left(\frac{\veps_b}{\rm 0.1 GeV} \right)^{\beta-2} - 
         \left(\frac{\veps_b}{\rm 10 GeV} \right)^{\beta-2}  \right]
\end{equation}
%with $f(\beta) = \ln 100$ for $\beta=2$.

\vspace*{2mm}
\subsection{Peak/break energy of the LAT component}

 A lower limit on the observer-frame break-energy $\veps_b$ can be set by requiring that the 
0.1--10 keV afterglow emission measured during the X-ray light-curve plateau (at 0.3-10 ks) by Swift/XRT, 
of about ${\cal F}_{xrt} \simeq 10^{-10}\fluxcgs$ (O'Brien et al 2006), is not dimmer than the 
extrapolation ${\cal F}_{lat}$ of the GeV afterglow spectrum, for an afterglow of GeV fluence 
$\Phi = 10^{-5} \ergcm2$ at $t=10$ s, decreasing as $\Phi \propto t^{-0.3}$, and with a high-energy 
slope $\beta = 1.1$:
\begin{equation}
 {\cal F}_{xrt} > {\cal F}_{lat}= 3.10^{-8} \left( \frac{t}{1\, {\rm ks}} \right)^{-1.3} 
   \left( \frac{\veps_b}{1\, {\rm keV}} \right)^{\alpha - 1.1} \, \fluxcgs
\end{equation}
For $\alpha = 2/3$, the high-energy spectral component does not overshine the X-ray plateau flux if 
$\veps_b (1\, ks) \simg 50$ keV. For $\alpha = 3/2$, the high-energy spectral component is dimmer 
than the X-ray plateau if $\veps_b (1\, ks) \simg 10$ MeV, but could be the X-ray plateau if 
$\veps_b \simeq 1 (t/1\, ks)^{-1.7 \pm 0.7}$ MeV.
This evolution is consistent with the $t^{-3/2}$ expected for the peak energy of the forward-shock 
synchrotron spectrum, but implies that $\veps_b (10 \,s) \simg 1$ GeV during the burst, which is 
inconsistent with LAT observations, that do not show a high-energy component peaking in the LAT
window.

 However, X-ray plateau measurements are often lacking during the early GeV afterglows monitored
by LAT, thus the above low limits on $\veps_b$ cannot be derived for individual afterglows.

 The sub-MeV burst light-curve may also set a constraint on the unknown $\veps_b$ in the following way.
For $\veps_b \simeq 10$ MeV and $\alpha=2/3$, the LAT spectral component yields a 100 keV flux 
${\cal F}_{lat} = (\Phi/t_{grb}) (0.1\, {\rm MeV}/\veps_b) ^{2-\alpha}= 2.10^{-9} \Phi_{-5} 
t_{grb,1}^{-1} \varepsilon_{b,7}^{-4/3} \fluxcgs$ during a $t_{grb} = 10$ s burst 
(using the notation $X_n = X(cgs)/10^n$ and measuring photon energies in eV).  This emission is 
sufficiently below the typical flux of a bright burst, ${\cal F}_{grb} = 10^{-5} \fluxcgs$, that 
it does not overshine a fast-decaying $F_{grb} \propto t^{-(2\div4)}$ tail (O'Brien et al 2006).
In contrast, an energy-break $\veps_b$ that falls below 100 keV will produce a burst emission 
${\cal F}_{lat} = (\Phi/t_{grb}) (0.1 {\rm MeV}/\veps_b)^{2-\beta}= 10^{-6} \Phi_{-5} t_{grb,1}^{-1} 
\fluxcgs$ (independent of $\veps_b$, for a high-energy LAT spectral slope $\beta \simeq 2$) that 
rivals that of the prompt emission.
 
 Thus, a bright LAT afterglow following a slowly-fading GRB {\sl may} have a break-energy $\veps_b 
< 100$ keV, but one following a burst with a steep decay {\sl must} satisfy $\veps_b \gg 1$ MeV 
during the burst tail.

%\vspace*{2mm}
\subsection{Optical thickness to photon-photon absorption}

\begin{figure*}
\centerline{\psfig{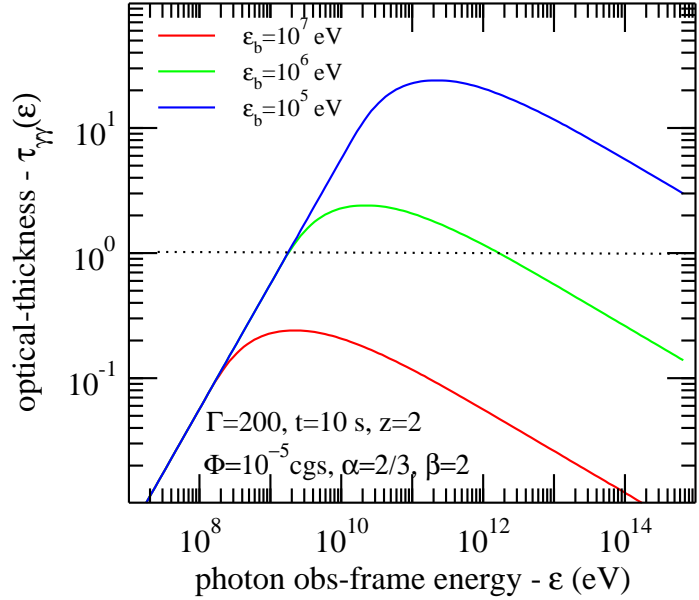}}
\figcaption{ Optical-thickness to pair-formation as function of observer-frame
   photon energy for a source of high-energy photons with the indicated parameters (fluence in cgs units)
   and for three values of the break energy $\veps_b$ .}
\end{figure*}
 For an isotropic distribution of photons in the frame of the shocked fluid, which moves at Lorentz factor 
$\Gamma$ in the lab-frame (at redshift $z$), the optical thickness to a photon of energy $\eps'_o$ is
\begin{equation}
 \taugg (\eps'_o) = \frac{\sigma_e}{4 \pi R^2} \int_0^\pi d\theta' \frac{\sin \theta'}{2}
                  \int_{\eps'_{th}}^\infty dN_\gamma(\eps') \fgg (\eps'_o\eps',\theta')
\label{taugg}
\end{equation}
where primed quantities are in the shock-frame, $\sigma_e$ is the Thomson cross-section for electron scattering,
\begin{equation}
 R (t) \simeq (z+1)^{-1} ct\Gamma^2
\label{R}
\end{equation}
is the source (shock) radius at observer-time $t$ (corresponding to the arrival-time of photons emitted
by the visible edge of the source; equality in equation above holds for an undecelerated source; a factor
4/3 applies to the right-hand side for a blast-wave decelerated by a wind-like medium), 
$\theta'$ is the angle of incidence between the test-photon
of energy $\eps'_o$ and a target-photon $\eps'$, $dP/d\theta' = (1/2) \sin \theta'$ is the probability of
two photons interacting at an angle $\theta'$
\begin{equation}
 \eps'_{th} (\eps'_o,\theta') = \frac{2 (m_e c^2)^2}{(1-\cos \theta') \eps'_o}
\label{ethr}
\end{equation}
is the threshold-energy for pair-formation, and
\begin{displaymath}
 \fgg (\eps'_o\eps',\theta') = \frac{3}{8x^2} \left[ \left( 2 + 2x^{-2} - x^{-4}\right) \ln \left(x + \sqrt{x^2-1} 
        \right) \right. 
\end{displaymath}
\begin{equation}
      - \left. \left( 1 + x^{-2} \right) \sqrt{1-x^{-2}}\right] = \frac{\sigma_{\gamma\gamma}}{\sigma_e} 
\label{fgg}
\end{equation}
is the cross-section for photon-photon absorption, with 
\begin{equation}
 x = \sqrt{\frac{1}{2} \frac{\eps'_o \eps'}{m_e^2 c^4} (1-\cos \theta')} = 
     \sqrt{ \frac{\eps'}{\eps'_{th}(\eps'_o,\theta')} } \geq 1 \; 
      {\rm for} \; \eps' \geq \eps'_{th}
\label{obs}
\end{equation}

 The integral in equation (\ref{taugg}) is calculated numerically; Figure 1 shows $\taugg (\veps)$ for
a photon of observer-frame energy 
\begin{equation}
 \veps = \frac{\eps}{z+1}=\frac{\Gamma \eps'}{z+1}
\label{zz}
\end{equation}
corresponding to the typical relativistic boost ($\Gamma$) of a photon of shock-frame energy $\eps'$.
To extract the dependence of $\taugg(\veps)$ on the source parameters $\Gamma$, $\Phi$, $z$ and observer
time $t$, an approximation to equation (\ref{taugg}) is needed. 
Extending the approximation $\fgg (x) = (3/4) \ln (2x)/x^2$, accurate for $x \gg 1$, to all $x \geq 1$, 
allows the second integral in equation (\ref{taugg}) to be calculated analytically, but the resulting
integral over the incidence angle $\theta'$ is not so nice. 
The dependence of $\taugg$ on source parameters can be obtained by setting $\fgg (x) = const$ and by
approximating 
\begin{equation}
 \int_{\eps'_{th}}^\infty dN_\gamma(\eps') \simeq \left.\frac{dN_\gamma}{d\eps}\right|_{\eps'_{th}} \eps'_{th}
\end{equation}
which leads to an integral over $\theta'$ that can be calculated easily. 
Dropping the integral over $\theta'$ and assuming that most pairs form at threshold are further
simplifications that lead to the correct dependence of $\taugg$ on source parameters:
\begin{equation}
 \taugg (\eps) \simeq \frac{\sigma_e}{4 \pi R^2} \left.\frac{dN_\gamma}{d\eps}\right|_{\eps_{th}} \eps_{th}(\eps)
\end{equation}
where 
\begin{equation}
 \eps_{th} (\eps) = 4 \Gamma^2 \frac{(m_e c^2)^2}{\eps}
\end{equation}
is the lab-frame threshold-energy for a lab-frame incidence angle $\theta = \Gamma^{-1} \ll 1$ (the source 
motion at $\Gamma$ collimates photons within an angle $\Gamma^{-1}$ around the direction of motion). 
Using equations (\ref{dNg}), (\ref{phi}), and (\ref{phi}), one arrives at
\begin{equation}
 \taugg (\veps) \propto \frac{\Phi}{t^2} \left\{  \begin{array}{ll}  
  \left(\frac{\displaystyle z+1}{\displaystyle \Gamma}\right)^{2\beta+2} \veps_b^{\beta-2} \veps^{\beta-1} 
   & \veps_b < \veps_{th}(\veps) \\
  \left(\frac{\displaystyle z+1}{\displaystyle \Gamma}\right)^{2\alpha+2} 
       \veps_b^{-(2-\alpha)} \veps^{-(1-\alpha)} & \veps_{th}(\veps) < \veps_b \\ \end{array}   \right.
\label{tau1}
\end{equation}
having switched to observer-frame photon energies (equation \ref{zz}). 
Using equation (\ref{ethr}), the conditions above become 
\begin{displaymath}
 \left\{ \begin{array}{lll} 
   \veps_b < \veps_{th}(\veps) & \rightarrow & \veps < \tveps \\
   \veps_{th}(\veps) < \veps_b & \rightarrow & \tveps < \veps \\
 \end{array}  \right.
\end{displaymath}
with $\tveps \equiv [4\Gamma^2 (m_e c^2)^2]/[(z+1)^2 \veps_b]=4.6\,\Z^{-2}\Gamma_{2.3}^2 \veps_{b,6}^{-1}$ 
GeV, where $\Z \equiv (z+1)/3$.

 Equation (\ref{tau1}) shows the obvious fact that $\taugg$ is proportional to the photon column density
($N_\gamma/R^2 \propto \Phi/t^2$), and that it has a strong dependence on the source Lorentz factor.
Given that $\alpha \geq 0$ and $\beta \simg 2$, equation (\ref{tau1}) also shows that $\taugg (\veps)$ increases
with photon energy for $\veps < \tveps$ and decreases with it for $\tveps < \veps$, with the maximal optical 
thickness reached at $\tveps$. 

 The coefficients missing in equation (\ref{tau1}) depend on the photon spectrum slopes $\alpha$ and $\beta$. 
Figure 1 shows optical thickness for the representative values $\alpha = 2/3$ and $\beta = 2$, obtained
numerically by integrating equation (\ref{taugg}). In the asymptotic power-law regimes, the numerical
approximation is
\begin{equation}
 \taugg (\veps) = \left\{  \begin{array}{ll}  
    0.57 \, \frac{\displaystyle \Z^6 \Phi_{-5}}{\displaystyle \Gamma_{2.3}^6 t_1^2} \,\veps_9 &  \veps < \tveps \\
  \\  2.6 \, \frac{\displaystyle \Z^{10/3} \Phi_{-5}}{\displaystyle \Gamma_{2.3}^{10/3} t_1^2 \veps_6^{4/3}} 
        \,\veps_{11}^{-1/3} &  \tveps < \veps \\
    \end{array}   \right.
\label{tau2}
\end{equation}
The two branches above intersect at
\begin{equation}
  \tveps \equiv 10 \, \Z^{-2} \frac{\Gamma_{2.3}^2}{\veps_{b,6}} \; {\rm GeV}
\label{etilde}
\end{equation}
where the optical thickness is maximal:
\begin{equation}
  \taugg(\tveps) = 5.7\, \frac{\Z^4 \Phi_{-5}}{\Gamma_{2.3}^4 t_1^2 \veps_{b,6}}
\label{taumax}
\end{equation}
Equation (\ref{tau2}) can now be written as
\begin{equation}
 \taugg (\veps) = \taugg (\tveps) \left\{ \begin{array}{ll}  
             \veps/\tveps  & \veps < \tveps \\ (\veps/\tveps)^{-1/3}  & \tveps < \veps \\ 
             \end{array}   \right.
\label{tau3}
\end{equation}
From here, it follows that 
\begin{equation}
 \taugg(\veps) < 1 \quad {\rm if} \quad \veps_b > \tveps_b \equiv 
    5.7\, \frac{\Z^4 \Phi_{-5}}{\Gamma_{2.3}^4 t_1^2} \, {\rm MeV}
\label{taulow}
\end{equation}
and
\begin{equation}
 \taugg(\veps) = \left\{ \begin{array}{ll}
  < 1 \;, & \veps < \veps_- \\ > 1 \;, & \veps_- < \veps < \veps_+ \\ < 1 \;, & \veps_+ < \veps \\ 
  \end{array} \right.  \quad {\rm if} \quad \veps_b < \tveps_b 
\label{taucomp}
\end{equation}
where
\begin{equation}
 \veps_- \equiv 1.75\, \frac{\Gamma_{2.3}^6 t_1^2}{\Z^6 \Phi_{-5}} \; {\rm GeV} \;, 
 \veps_+ \equiv 1.82\, \frac{\Z^{10}\Phi_{-5}^3}{\Gamma_{2.3}^{10} t_1^6 \veps_{b,6}^4} \; {\rm TeV} 
\label{epm}
\end{equation}
For $\beta = 2$, the optical-thickness to photon-photon absorption is independent of $\veps_b$ for
photons of energy lower than $\tveps$ (i.e. photons with threshold energy above $\veps_b$), hence
the lower limit $\veps_-$ above which the photon-front is optically-thick is also independent of
$\veps_b$, as illustrated in Figure 1.

 Equation (\ref{taulow}) can be reinterpreted as 
\begin{equation}
 \taugg(\veps) < 1 \quad {\rm if} \quad \Gamma > \Gamma_\pm \equiv 
    310\, \frac{\Z \Phi_{-5}^{1/4}}{t_1^{1/2}\veps_{b,6}^{1/4}}
\end{equation}
For $\Gamma < \Gamma_\pm$, the photon-front is optically-thick to photons in the $(\veps_-,\veps_+)$
range, which widens with decreasing $\Gamma$ (see equation \ref{epm}), covering the entire LAT window
if $\veps_- \siml 100$ MeV, which is equivalent to $\Gamma \siml \Gamma_{lat} \equiv 125\, \Z \Phi_{-5}^{1/6} 
t_1^{-1/3}$. For such low Lorentz factors, the LAT emission is heavily absorbed and the afterglow high-energy
emission undetectable. At the other extreme, if $\Gamma \simg 2.2\, \Gamma_{lat}$, then $\veps_- \simg 10$ GeV 
and the LAT emission is weakly absorbed. For $\Gamma_{lat} < \Gamma < 2.2\, \Gamma_{lat}$, the LAT emission
is moderately absorbed, photon-photon absorption rendering a spectrum that curves downward at higher
energies, for a power-law intrinsic spectrum (e.g. figure 2 of Panaitescu et al 2014). 

 Thus, perfect power-law LAT spectra set only a lower limit on the source Lorentz factor: $\Gamma \simg 300\,
\Phi_{-5}^{1/6} t_1^{-1/3}$. Such a weak dependence on the afterglow fluence $\Phi$ and epoch $t$ of 
observations suggests that the measurement of curvature in the LAT spectrum would yield a fairly accurate 
determination of $\Gamma$. Obviously, the non-detection of the high-energy afterglow emission is not 
necessarily proof of high absorption and does not tell us anything about $\Gamma$.

\vspace*{2mm}
\subsection{Total number of pairs}
\label{number}
 
 The total number of pairs is an integral over the photon spectrum of the absorbed fraction 
$g[\taugg(\veps)$
\begin{equation}
 N_\pm = \int_0^\infty d\veps \frac{dN_\gamma}{d\veps} g[\taugg(\veps)]
\label{N0}
\end{equation}
with the photon spectrum of equation (\ref{dNg}).
 To calculate the fraction of absorbed photons corresponding to the optical-thickness to pair-formation 
$\taugg(\veps)$ (equation \ref{tau2}), consider a medium of geometrical thickness $\Delta$ and linear 
absorption coefficient $\alpha$, and in which the production and absorption of photons is homogeneous
(same at any location).Then, the fraction of photons that are absorbed is
\begin{equation}
  g = \int_0^\Delta \frac{dx}{\Delta} (1 - e^{-\alpha (\Delta - x)}) = 1 - \frac{1-e^{-\tau}}{\tau}
    \simeq \left\{ \begin{array}{ll}  \tau/2 & \tau \ll 1 \\ 1 & \tau \gg 1 \\ \end{array} \right.
\label{g}
\end{equation}
obtained by integrating the photon absorption from the medium inner edge ($x=0$) to its outer boundary 
($x=\Delta$), and with $\tau = \alpha \Delta$.  

 In the case of pair production in a decelerating source, the photons radial distribution is not uniform.
In this case, the fraction of absorbed photons is
\begin{equation}
  g = \int_0^1 dy \frac{dn_\gamma}{dy}(1 - e^{-\tau(y \rightarrow 1)}) \;,
\label{gg}
\end{equation}
where $dn_\gamma/dy$ is radial distribution of photons normalized by $\int_0^1 dy (dn_\gamma/dy) =1$ and
\begin{equation}
  \tau (y \rightarrow 1) = \int_y^1 \alpha(z) dz = \tau (0 \rightarrow 1) \int_y^1 \frac{dn_\gamma}{dz} dz
\end{equation}
is the absorption optical thickness from coordinate $y = x/\Delta$ to the outer edge at $y=1$,
$\tau (0 \rightarrow 1) \equiv \tau$ being the entire optical thickness of the medium.
Substituting in equation (\ref{gg}), we get 
\begin{displaymath}
  g =  \int_0^1 \frac{dn_\gamma}{dy} dy - 
   \int_0^1 dy \frac{dn_\gamma}{dy} \exp\left( -\tau \int_y^1 \frac{dn_\gamma}{dz} dz \right)
\end{displaymath}
\begin{equation}
    = 1 - \frac{1}{\tau} \int_0^1 dy \frac{d}{dy} \left[ 
     \exp\left( -\tau \int_y^1 \frac{dn_\gamma}{dz} dz \right) \right]
\end{equation}
\begin{displaymath}
   = 1 - \frac{1}{\tau} \left[ 1- \exp\left( -\tau \int_0^1 \frac{dn_\gamma}{dy} dy \right) \right]
   = 1 - \frac{1-e^{-\tau}}{\tau}
\end{displaymath}
Therefore, as long as the absorption coefficient $\alpha$ is proportional to the density of the 
to-be-absorbed photons, the fraction of absorbed photons from a medium depends only on the optical 
thickness $\tau = \int \alpha(z) dz$ of that medium and does not "care" about the exact spatial 
variation of $\alpha$.

 The integral in equation (\ref{N0}) is calculated numerically; for an analytical estimate, we use the 
approximation given in equation (\ref{g}).

 {\bf Case 1.} 
For $\veps_b > \tveps_b/2$ (equation \ref{taulow}), the maximal optical thickness (equation \ref{taumax})
satisfies $\taugg(\tveps) < 2$, hence $\taugg (\veps) < 2$ for any photon. With two branches for the photon
spectrum (equation \ref{dNg}) and two for the optical thickness (equation \ref{tau1}), the integral in equation 
(\ref{N0}) splits in three integrals  
\begin{displaymath} 
 N_\pm = \int_\veps dN_\gamma \frac{\taugg(\veps)}{2} 
\end{displaymath} 
\begin{equation} 
  = \frac{1}{2} \left.\frac{dN_\gamma}{d\veps}\right|_{\veps_b} \taugg (\tveps) 
 \left[ \int_0^{\veps_b} d\veps \left(\frac{\veps}{\veps_b}\right)^{-\alpha} 
   \left(\frac{\veps}{\tveps}\right)^{\beta-1} + \right.
\label{int1}
\end{equation}
\begin{displaymath} 
 \left.  + \int_{\veps_b}^{\tveps} d\veps \left(\frac{\veps}{\veps_b}\right)^{-\beta} 
         \left(\frac{\veps}{\tveps}\right)^{\beta-1} 
   + \int_{\tveps}^{\infty} d\veps \left(\frac{\veps}{\veps_b}\right)^{-\beta} 
    \left(\frac{\veps}{\tveps}\right)^{\alpha-1}  \right]
\end{displaymath} 
for the more likely case $\veps_b < \tveps$. Using equation (\ref{taumax}), this condition requires 
that $\veps_b < 100\, \Z^{-1}\Gamma_{2.3}$ MeV, which is satisfied by LAT spectra and which, 
together with the working condition $\veps_b > \tveps_b/2$, requires that $\Gamma > \tGamma$ where
\begin{equation} 
 \tGamma \equiv 100\, \Z \frac{\Phi_{-5}^{1/5}}{t_1^{2/5}} 
\label{tGam}
\end{equation}
The scaling of the integrals in equation (\ref{int1}) with $\veps$ is $\veps^{\beta-\alpha}$,
$\ln \veps$, and $\veps^{-(\beta-\alpha)}$, respectively; taking into account that $\beta > \alpha$, 
this implies that most pairs are formed from (the second integral, corresponding to) photons with 
energy above the spectral break $\veps_b > \tveps_b/2 \sim$ few MeV and below the energy for maximal 
optical thickness $\tveps \siml$ 10 GeV, interacting with photons above threshold energies of about 
1 GeV and 1 MeV, respectively.

 For $\alpha = 2/3$ and $\beta = 2$, equation (\ref{int1}) yields
\begin{equation} 
 N_\pm \stackrel{2\veps_b > \tveps_b}{=} 10^{55} \left( 0.67 + 0.22 \ln \frac{\Gamma_{2.3}}{\Z\veps_{b,7}} 
      \right) \Z^8 \frac{\Phi_{-5}^2}{t_1^2 \Gamma_{2.3}^6}
\label{N1}
\end{equation}
Using equation (\ref{R}), the optical-thickness to photon scattering of the pairs is
\begin{equation} 
 \tau_\pm = \frac{2 \sigma_e N_\pm}{4 \pi R^2} = 0.024 \left( 1 + 0.33 \ln \frac{\Gamma_{2.3}}{\Z\veps_{b,7}} 
      \right) \Z^{10} \frac{\Phi_{-5}^2}{t_1^4 \Gamma_{2.3}^{10}}
\label{taupairs}
\end{equation}
Ignoring the logarithmic term, this implies that the pairs are optically thin (to photon scattering in the
Thomson regime, because most pairs are cold) for $\Gamma > \Gamma_\tau$ with
\begin{equation} 
  \Gamma_\tau \equiv 138\, \Z \Phi_{-5}^{1/5} t_1^{-2/5} 
\label{Gtau}
\end{equation}

 {\bf Case 2.} 
For $\veps_b < \tveps_b/2$, $\taugg(\veps)$ is relative to 2 as in equation (\ref{taucomp}) but with
$\veps_-$ larger by a factor $2^{1/(\beta-1)}$ than in equation (\ref{epm}) and $\veps_+$ smaller by a 
factor $2^{1/(1-\alpha)}$ than in (\ref{epm}). Having two branches for the photon spectrum and three for 
the optical thickness, the integral of equation (\ref{N0}) splits in four:
\begin{displaymath} 
 N_\pm = \int_\veps dN_\gamma \min \left\{\frac{\taugg(\veps)}{2},1\right\} 
  = \left. \frac{dN_\gamma}{d\veps}\right|_{\veps_b} \times
\end{displaymath} 
\begin{displaymath} 
 \left[ \frac{1}{2}\taugg (\tveps) \int_0^{\veps_b} d\veps \left(\frac{\veps}{\veps_b}\right)^{-\alpha} 
    \left( \frac{\veps}{\tveps} \right)^{\beta -1}+ \right.
\end{displaymath} 
\begin{equation} 
   \frac{1}{2} \taugg (\tveps) \int_{\veps_b}^{\veps_-} d\veps \left(\frac{\veps}{\veps_b}\right)^{-\beta} 
    \left( \frac{\veps}{\tveps} \right)^{\beta -1} + 
\label{int2}
\end{equation} 
\begin{displaymath} 
   \int_{\veps_-}^{\veps_+} d\veps \left(\frac{\veps}{\veps_b}\right)^{-\beta} +
   \left. \frac{1}{2} \taugg (\tveps) \int_{\veps_+}^{\infty} d\veps \left(\frac{\veps}{\veps_b}\right)^{-\beta} 
    \left( \frac{\veps}{\tveps} \right)^{\alpha -1}  \right]
\end{displaymath} 
for the more likely case $\veps_b < \veps_-$ (for $\Gamma > \tGamma$, this is implied by the working
condition $\veps_b < \tveps_b/2$).  

 The integrals in equation (\ref{int2}) show that most pairs form from (the second integral, corresponding 
to) photons above the spectral break $\veps_b < \tveps_b/2 \sim 3$ MeV and below $\veps_- \sim 1$ GeV, 
for which the photon front is optically thin, interacting with photons above threshold energies $> 1$ GeV 
and few MeV, respectively. 
%Thus, in general, most pairs are produced by high-energy photons for which the front is 
%optically thin to pair-formation.

 For $\alpha = 2/3$ and $\beta =2$, equation (\ref{int2}) leads to
\begin{equation} 
 N_\pm \stackrel{2\veps_b < \tveps_b}{=} 10^{55} \left( 1.1 + 0.11 \ln \frac{\Gamma_{2.3}^6 t_1^2}{\Z^6 \Phi_{5} 
   \veps_{b,6}} \right) \Z^8 \frac{\Phi_{-5}^2}{t_1^2 \Gamma_{2.3}^6}
\label{N2}
\end{equation}
After calculating the pair optical thickness to photon scattering as done above for $\veps_b > \tveps_b/2$,
it can be shown that the minimal Lorentz factor for optical-thinness is close to that in equation (\ref{Gtau}).

 Equations (\ref{N1}) and (\ref{N2}) show that the number of pairs formed has a weak dependence on the
(unknown) break $\veps_b$ of the photon spectrum, varies like $\Phi^2$ (as expected for a two-photon 
interaction), and has a strong dependence on the source Lorentz factor, resulting in part from the 
dependence of the threshold energy for pair formation on $\Gamma$ and in part from the decrease
of the photon density with source radius (which is proportional to $\Gamma^2$).

\vspace*{2mm}
\subsection{Scattering of afterglow photons on internal leptons}

 It is worth comparing the number of pairs with that of electrons existing in the two possible
source of GeV afterglow photons, the forward and reverse shocks.

 For a Wolf-Rayet GRB progenitor with a mass-loss rate $dM/dt = 10^{-5}\, M_\odot/{\rm yr}$, blowing a wind
of terminal velocity $v_w=10^3 \; {\rm km/s}$, the wind baryon density is
\begin{equation}
 n(R) = \frac{dM/dt}{4\pi m_p v_w R^2} = \frac{3.0 \times 10^{35}}{R^2} \; {\rm cm^{-3}}
\label{WR}
\end{equation}
The ratio of the number of formed leptons $N_l = 2 N_\pm$ (eqs \ref{N1} and \ref{N2}) to the electrons 
energized by the forward-shock is
\begin{equation}
 \frac{N_l}{N_{fs}} = \frac{N_\pm}{\pi n R^3} \simeq 2600 \frac{\Z^9 \Phi_{-5}^2}{t_1^3 \Gamma_{2.3}^8}
\label{ratio}
\end{equation}
taking into account that the wind-like medium is made of elements heavier than hydrogen (with one 
electron for two baryons). Thus, for $\Gamma < \Gamma_{fs} \equiv 540\, \Z^{9/8} \Phi_{-5}^{1/4} t_1^{-3/8}$, 
the pairs are more numerous than the forward-shock electrons. 

 If we assume that $i)$ the reverse and forward shock baryons contain about the same (kinetic plus thermal) 
energy and $ii)$ the ejecta are normal matter (with one electron per baryon), then the number of ejecta 
electrons is at most $\Gamma$ times larger than that of the forward-shock's (as in the case of a 
semi-relativistic reverse-shock): $N_{rs} \siml \Gamma N_{fs}$. Then, the number of pairs exceeds 
that of the ejecta electrons for $\Gamma \siml \Gamma_{rs} \equiv 270\, \Z \Phi_{-5}^{2/9} t_1^{-1/3}$. 

 The above suggest that emission from pairs is of importance for GeV afterglow sources with a Lorentz
factor in the few hundreds, but pairs may radiate at a different energy than the reverse or forward-shock
electrons, where the pairs could dominate the afterglow emission even if they are fewer.

% Scattering of seed photons with source-frame energy $\eps' > 1$ MeV on {\sl hot} leptons occurs in 
%the Klein-Nishina regime (with a reduced cross-section and scattering optical thickness) decollimates 
%slightly (in the lab-frame) the $\Gamma^{-1}$ cone of seed photons, and increases substantially the
%energy of the scattered photon, hence its pair-formation threshold energy is lower and the probability
%that it forms a pair while crossing the photon front is higher, provided that $\taugg(\eps') < 1$. 
%Thus, scattering on hot leptons could increase $N_\pm$, the number of  pairs.

 It is also worth investigating if scattering of pair-forming photons on existing (reverse and 
forward-shock) electrons or on the already-formed pairs could change significantly the number of pairs
formed from unscattered photons. Most photon scattering occurs on leptons that are cold. That is 
certainly the case for the pairs, most of which are born cold (see the distribution of formed 
pairs with energy in Figure 2, left panel), and is likely true for the (reverse-shock) ejecta electrons 
and the ambient medium electrons (swept-up by the forward-shock) because they should be cooling fast 
radiatively if their synchrotron emission were to account for the observed GeV afterglow. 

 For a scattering optical thickness $\tau_{sc}$, the effective photon-photon attenuation thickness
is $\taugg = \sqrt{\taugg(\taugg + \tau_{sc})}$, therefore scattering on cold leptons is negligible 
when $\taugg (\veps) > \tau_{sc}$. As shown in \S\ref{number}, most pairs form from photons with energy
$\veps < \tveps$, for which $\taugg (\veps)$ is that in the first branch of equation (\ref{tau2}). 
The optical thickness to Thomson scattering on reverse-shock electrons (which are more numerous than 
in the forward-shock) is $\tau_{sc} \siml \sigma_e (\Gamma N_{fs})/(4 \pi R^2) \siml 5.10^{-3} 
\Z/(\Gamma_{2.3} t_1)$. Thus, $\taugg (\veps) > \tau_{sc}$ is satisfied for 
$\veps \simg \veps_{rs} \equiv 10 (\Gamma_{2.3}/\Z)^5 (t_1/\Phi_{-5})$ MeV. 

 The optical thickness $\tau_\pm$ to Thomson scattering on (already formed) pairs is that given in equation 
(\ref{taupairs}), thus $\taugg(\veps) > \tau_\pm$ is satisfied for $\veps > \veps_{pair}^{(Th)} \equiv 50 
(\Z/\Gamma_{2.3})^4 (\Phi_{-5}/t_1^2)$ MeV. Scattering on cold electrons of observer-frame photons with
energy above $\veps_{kn} \equiv \Gamma m_e c^2/(z+1) = 34 \Z^{-1} \Gamma_{2.3}$ MeV occurs in the 
Klein-Nishina regime. Given that $\veps_{pair}^{(Th)} \simg \veps_{kn}$, it is worth considering scattering 
on pairs in the KN regime, when the scattering cross-section $\sigma_{kn} (\veps) \simeq (3/8) \sigma_e 
\ln(2x)/x$ with $x = \veps/\veps_{kn}$. In this case, the condition $\taugg (\veps) > \tau_\pm (\veps)$ 
is satisfied for $\veps > \veps_{pair}^{(kn)} \equiv 25 (\Z/\Gamma_{2.3})^{3/2} (\Phi_{-5}^{1/2}/t_1)$ MeV. 

 To the above identification of the photon energies for which scattering increases significantly the photon 
escape path and attenuation, we add that, according to equations (\ref{int1}) and (\ref{int2}), most pairs 
are formed from photons with energy in the range $(\veps_b,\tveps)$ (for $\veps_b > \tveps_b/2$) or 
$(\veps_b,\veps_-)$ (for $\veps_b < \tveps_b/2$). This means that scattering is important for photons in 
the lower part of those energy intervals and not so important in the upper part. 
Equations (\ref{int1}) and (\ref{int2}) also show that each decade of photon energy provides an equal 
contribution to the number of pairs. Therefore, scattering has a negligible effect on the number of pairs 
if the logarithmic length of the upper interval is larger than that of the lower interval. 
Using the expressions for $\tveps$, $\veps_-$, $\veps_{rs}$, and $\veps_{pair}^{(kn)}$,
it can be shown that scattering on pairs should not increase much the number of pairs if
$\Gamma > 120$ (for $\veps_b > \tveps_b$) and $\Gamma > 140$ (for $\veps_b < \tveps_b$), while
scattering on reverse-shock electrons is negligible if $\Gamma < 370$ (for $\veps_b > \tveps_b$) and 
$\Gamma < 780$ (for $\veps_b < \tveps_b$), having left out sub-unity powers of the parameters $\Z,\Phi,t$. 
Adding to these that the reverse-shock electrons are more numerous than the pairs if $\Gamma > \Gamma_{rs}
= 270$, it follows that scattering of the pair-forming photons below the LAT range increases the total 
number of pairs only through scattering by reverse-shock electrons and only if $\Gamma$ is at least several 
hundreds.

\vspace*{2mm}
\subsection{Pair distribution with energy}

 The shock-frame energy of a pair $\eps'_p$ depends on the energies of the incident photons, test-photon
 $\eps'_o$ and target-photon $\eps'$, the incidence angle $\theta'$, and the center-of-momentum (CoM) 
frame angle $\phi''$ at which the electron and positron emerge, measured relative to the direction of 
motion of the photons' CoM. In the shocked-fluid frame, the CoM moves at velocity $\vec{\beta'}_{cm} = 
(\vec{\eps'}_o + \vec{\eps'})/ (\eps'_o+\eps')$, the corresponding Lorentz factor being
\begin{equation} 
 \Gamma'_{cm} = \frac{\eps'_o +\eps'}{\sqrt{2 \eps'_o \eps' (1-\cos \theta')}}  
\end{equation}
In the CoM frame, the incident photons have the same energy 
\begin{equation} 
 \eps'' = \left[ \frac{1}{2} \eps'_o \eps' (1-\cos \theta') \right]^{1/2} = 
  \left[ \frac{\eps'}{\eps'_{th}(\eps'_o,\theta')} \right]^{1/2}
\end{equation}
collide head-on, and form an electron and a positron of equal energy $\eps''$, moving in opposite 
directions, at angles $\phi''$ and $\pi - \phi''$ relative to $\vec{\beta'}_{cm}$. Then, the shock-frame 
electron and positron energies are
\begin{equation} 
 \eps'_\pm = \Gamma'_{cm} (\eps'' \pm p''c \beta'_{cm} \cos \phi'')
\label{epm1}
\end{equation}
where $p''c = \sqrt{\eps''^2 - m_e^2 c^4}$ is the electron/positron momentum in the CoM frame.

 To obtain the distribution of formed leptons with their shock-frame energy, $dN_l/d\eps'_\pm$, 
we start from equation (\ref{N0}) in shock-frame photon energy $\eps_o'$
\begin{equation} 
 \frac{dN_l}{d\eps'_o} = 2 \frac{dN_\gamma}{d\eps'_o} g[\taugg(\eps'_o)]
    = \frac{dN_\gamma}{d\eps'_o}  \left\{ \begin{array}{ll}  
    \taugg(\eps'_o) & \taugg(\eps'_o) \ll 1 \\ 2 & \taugg(\eps'_o) \gg 1 \\ \end{array} \right.
\label{cici}
\end{equation}
where the factor 2 accounts for two leptons being created from one photon.
The term $\taugg(\eps'_o)$ of the first branch offers a way to calculate $dN_l/d\eps'_\pm$, by expanding 
it (as in equation \ref{taugg}), leading to the differential pair-number in the 4-dimensional space 
$[\eps'_o,\eps',\theta',\phi'']$ 
\begin{equation} 
 \frac{d^4N_l}{d\eps'_o d\Omega' d\eps' d\Omega''} = 
   \frac{\sigma_e}{4\pi R^2} \frac{dN_\gamma}{d\eps'_o} \frac{dP}{d\Omega'} 
   \frac{dN_\gamma}{d\eps'} \frac{dP}{d\Omega''} \fgg (\eps'_o\eps',\theta') 
\label{num0}
\end{equation}
where $dP/d\Omega'' = \sin \phi''/2$ (because pairs emerge isotropically in the CoM frame), with $\fgg$ 
the photon-photon absorption cross-section of equation (\ref{fgg}). 
 For an isotropic distribution of the incident photons ($dP/d\Omega' = \sin \theta'/2$) 
\begin{equation} 
 \frac{d^4N_l}{d\eps'_o d\theta' d\eps' d(\cos \phi'')} = 
   \frac{\sigma_e}{16\pi R^2} \frac{dN_\gamma}{d\eps'_o}  \sin \theta'
   \frac{dN_\gamma}{d\eps'} \fgg (\eps'_o\eps',\theta') 
\label{num1}
\end{equation}
From equation (\ref{epm1}), $d\eps'_\pm = \pm \Gamma'_{cm}\beta'_{cm}p''c (d\cos \phi'')$, thus
\begin{displaymath} 
 \frac{d^4N_l}{d\eps'_o d\theta' d\eps' d\eps'_\pm} =
 \frac{d^4N_l}{d\eps'_o d\theta' d\eps' d(\cos \phi'')} \left| \frac{d\cos \phi''}{d\eps'_\pm} \right|
\end{displaymath}
\begin{equation} 
  = \frac{\sigma_e}{16\pi R^2} \frac{dN_\gamma}{d\eps'_o}  \sin \theta'
   \frac{dN_\gamma}{d\eps'} \frac{\fgg (\eps'_o\eps',\theta')}{\Gamma'_{cm}\beta'_{cm}p''c} 
\end{equation}
Then, the distribution of leptons with energy is the integral of the differential pair-number above
over the energies of the incident photons and over the incidence angle 
\begin{displaymath} 
 \frac{dN_l}{d\eps'_\pm} = \frac{\sigma_e}{16\pi R^2} 
   \int_0^\infty d\eps'_o \frac{dN_\gamma}{d\eps'_o} \int_0^\pi d\theta' \sin \theta'
\end{displaymath}
\begin{equation} 
   \int_{\eps'_{min}}^{\eps'_{max}} d\eps' \frac{dN_\gamma}{d\eps'} 
   \frac{\fgg (\eps'_o\eps',\theta')}{\sqrt{(\Gamma_{cm}^{'2} - 1)(\eps''^2 - m_e^2 c^4)}} 
\label{num2}
\end{equation}
where the limits $\eps'_{min} (\eps'_\pm)$ and $\eps'_{max} (\eps'_\pm)$ on the integral over the
spectrum of target photons are determined from
\begin{equation} 
 \Gamma'_{cm} (\eps'' - p''c \beta'_{cm}) \leq \eps'_\pm \leq \Gamma'_{cm} (\eps'' + p''c \beta'_{cm})
\end{equation}
together with $\eps' \geq \eps'_{th} (\eps'_o,\theta')$.
Unfortunately, the term $\Gamma'_{cm}\beta'_{cm}p''$ leads to a fourth degree equation in $\eps'$ that 
cannot be solved analytically to obtain the integral limits $\eps'_{min} (\eps'_\pm)$ and $\eps'_{max} 
(\eps'_\pm)$. 

\begin{figure*}
\centerline{\psfig{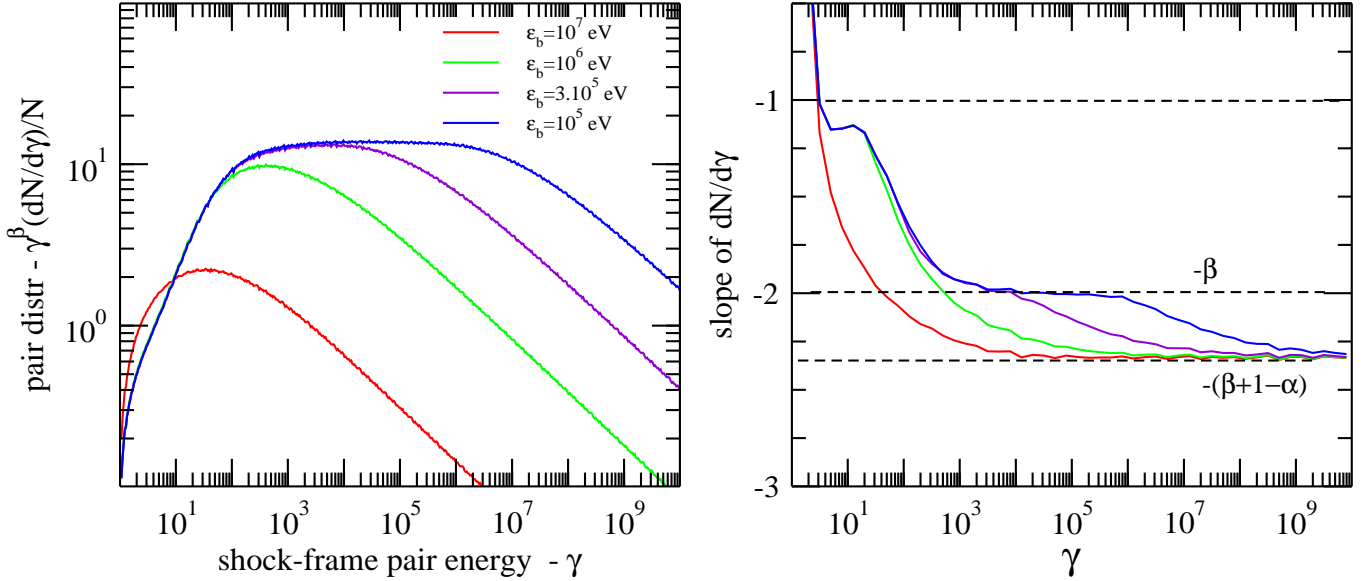}}
\figcaption{ Left panel: distribution with energy (multiplied by $\gamma^\beta$, to show where
   the power-law $dN_l/d\gamma \propto \gamma^{-\beta}$ branch occurs) of the pairs formed in a 
   relativistic source with same parameters as for Fig 1, calculated numerically by integrating equation 
   (\ref{num1}).
   Right panel: comparison between the derivative $d\log (dN_l/d\gamma)/d\log \gamma$ of the
   numerical pair distribution and the expected power-law slopes (equations \ref{dN1} and \ref{dN2}). 
   For $\veps_b > \tveps_b$ (red line), the front is optically thin for any photon and the lowest-energy 
   expected branch is not developed. For $\veps_b \ll \tveps_b$, all three branches are seen.}
%fig2
\end{figure*}

 Those limits can be calculated numerically and used to integrate equation (\ref{num2}), with the
following two corrections. First, a multiplicative factor $g[\taugg(\eps'_o)]/ \taugg(\eps'_o)$ 
should be applied to the $\eps'_o$ integrand, to account for the correct absorption fraction 
$g(\veps'_o)$. That ensures that no more than $dN_\gamma (\eps'_o)$ photons form pairs. Second,
a multiplicative factor $\min \{1, N_\gamma [>\eps'_{th}(\eps'_o)]/[N_\gamma (>\eps'_o) 
g(\taugg(\eps'_o))] \}$ ensures that the number of absorbed test photons $N_\gamma (>\eps'_o) 
g[\taugg(\eps'_o)]$ does not exceed the number of target photons $N_\gamma [>\eps'_{th}(\eps'_o)]$
above the threshold for pair-formation.

 The slopes of the pair distribution with energy $dN_l/d\eps'_\pm$ that results from 
integrating equation (\ref{num1}) can be inferred if the crude approximation $\eps'_\pm \simeq \eps'_o/2$
is made. This approximation is suggested by that pairs emerge most likely at an CoM-frame angle 
$\phi'' = \pi/2$, hence $\eps'_\pm = \Gamma'_{cm} \eps'' = (\eps'_o + \eps')/2$ (from equation 
\ref{epm1}), and by that most pairs are formed from a test-photon of energy $\eps'_o$ larger than 
the $\eps'$ of the target-photon. The latter is suggested by that most pairs are formed in interactions 
with target-photons close to (but above) the threshold for pair-formation (the integrand in equation 
\ref{taugg} shows that $\taugg \propto \eps' (dN_\gamma/d\eps') \fgg (\eps') \propto (\eps')^{1-\alpha} 
\ln (2x)/x^2$ with $x^2 \propto \eps'_o \eps'$, thus $\taugg \propto (\eps')^{-\alpha}$ with $\alpha > 0$) 
and by that $\eps'_o \gg \eps'_{th}(\eps'_o)$ (optical thickness $\taugg(\veps)$ is maximal at photon 
energy $\tveps \gg \veps_b = \veps_{th} (\tveps)$ -- equation \ref{taumax}).

 The approximation $\eps'_\pm \simeq \eps'_o/2$ implies that $dN_l/d\eps'_\pm \propto dN_\gamma/d\eps'_\pm$. 
Then, equation (\ref{cici}) leads to
\begin{equation}
 \frac{dN_l}{d\eps'_\pm} \propto \frac{dN_\gamma}{d\eps'_\pm} g[\taugg(\eps'_\pm)] \simeq 
   \frac{dN_\gamma}{d\eps'_\pm} \min \left\{1,\taugg(\eps'_\pm)\right\}
\end{equation}
with an approximation for the absorption factor $g$ that has the correct dependence on $\taugg$.
From here, the slopes of $dN_l/d\eps'_\pm$ can be easily calculated using the photon spectrum
(equation \ref{dNg}) and the optical thickness (equation \ref{tau1}).

 For $\tveps_b < \veps_b$, we have $\taugg(\veps) < 1$ for any photon, thus
\begin{equation}
 \frac{dN_l}{d\gamma} \propto \left\{ \begin{array}{ll} 
    \gamma^{-1}                & \eps'_b < \gamma m_ec^2 < \teps' \\
    \gamma^{-(\beta+1-\alpha)} & \teps' < \gamma m_ec^2 \\
   \end{array} \right.
\label{dN1}
\end{equation}
with $\teps'$ the shock-frame photon energy for which $\taugg$ is maximal (given in equation 
\ref{taumax}), $\gamma = \eps'_\pm/(m_ec^2)$ the pair's random Lorentz factor in the shock-frame and 
$\eps' = \eps (z+1)/\Gamma$ the shock-frame photon energy corresponding to the observer-frame $\veps$.
A $\gamma m_e c^2 < \eps'_b$ branch does not exist because $\eps'_o < \eps'_b$ photons form pairs in 
interaction with photons of energy $\eps' > \eps'_b$ and the corresponding pair energy $\eps'_\pm
\simeq \eps'/2$ is in one of the branches above.
The above distribution was derived for $\eps'_b < \teps'$ (which requires $\Gamma > \tGamma$)
but is also correct for $\eps'_b > \teps'$ (requiring that $\Gamma > \tGamma$) with $\eps'_b$
and $\teps'$ swapped.

 For $\veps_b < \tveps_b$, we have $\taugg(\veps) > 1$ for $\veps_- < \veps < \veps_+$ and
$\taugg(\veps) < 1$ otherwise, thus
\begin{equation}
 \frac{dN_l}{d\gamma} \propto \left\{ \begin{array}{ll} 
    \gamma^{-1}                & \eps'_b < \gamma m_ec^2 < \eps'_- \\
    \gamma^{-\beta}            & \eps'_- < \gamma m_ec^2 < \eps'_+ \\
    \gamma^{-(\beta+1-\alpha)} & \eps'_+ < \gamma m_ec^2 \\
   \end{array} \right.
\label{dN2}
\end{equation}
for the more likely case $\eps'_b < \eps'_-$, i.e. for $\Gamma > \tGamma$ (equation \ref{tGam}). 
If $0.63\, \tGamma < \Gamma < \tGamma$, then $\eps'_- < \eps'_b < \eps'_+$ and the first branch
above is $\gamma^{-\alpha}$ with $\eps'_b$ and $\eps'_-$ swapped. For $\Gamma < 0.63\, \tGamma$, 
we have $\eps'_+ < \eps'_b$ and, in addition to the preceding case, the second branch above is
$\gamma^{-1}$, with $\eps'_b$ and $\eps'_+$ swapped.

 Equations (\ref{dN1}) and (\ref{dN2}) indicate that the pair distribution with energy has up to
three power-law branches, with four possible values ($-1, -\alpha, -\beta, -\beta -1 +\alpha$) 
for the slope of each branch. Figure 2 shows that the pair distribution obtained numerically by 
integrating equation (\ref{num1}) displays only the highest-energy branch for $\tveps_b  < \veps_b$ 
because the range of energies over which the lowest-energy branch occurs is too narrow, from 
$\gamma = 1$ to
\begin{equation}
 \gamma (\teps') = \frac{(z+1)\tveps}{2\Gamma m_ec^2} = 15\, \frac{\Gamma_{2.3}}{\Z \veps_{b,7}} 
\end{equation}
For $\veps_b \ll \tveps_b$, all three branches are found in the numerical pair-distribution. 
The lowest-energy branch is short, extending from $\gamma \sim$ few to 
\begin{equation}
 \gamma (\eps'_-) = \frac{(z+1)\veps_-}{2\Gamma m_ec^2} = 26\, \frac{\Gamma_{2.3}^5 t_1^2}{\Z^5 \Phi_{-5}}. 
\end{equation}
while the second branch extends from $\gamma (\eps'_-)$ to 
\begin{equation}
 \gamma (\eps'_+) = \frac{(z+1)\veps_+}{2\Gamma m_ec^2} = 10^5 \frac{\Z^{11} \Phi_{-5}^3}
   {\Gamma_{2.3}^{11} t_1^2  \veps_{b,6}^4}
\end{equation}

 The largest photon energy measured by LAT, $\veps \simeq 100$ GeV, corresponds to a pair Lorentz
factor $\gamma \simeq 1500 \Z \veps_{11}/\Gamma_{2.3}$, which is between $\gamma(\eps'_-)$ and 
$\gamma(\eps'_+)$ (because $\veps_- < 100\, GeV < \veps_+$). In further calculations, we make the
assumption that the LAT power-law spectrum extends well above 100 GeV, implying that the power-law
distribution of pairs extends to $\gamma > 10^3$, as in Figure 2. A cut-off in the photon spectrum
above 100 GeV does not affect much the total number of pairs formed but reduces the number of higher 
energy pairs and, consequently, the optical flux from pairs formed in-between shocks.

\vspace*{2mm}
\subsection{Evolution of pair distribution}

 The evolution (with observer time $t$) of the leptons distribution with energy, $\N (\gamma) 
\equiv dN/d\gamma$, is given by
\begin{equation}
  \frac{\partial \N}{\partial t} = \N_{inj} +  \frac{\partial}{\partial \gamma} 
     \left( \N \left| \frac{d\gamma}{dt}\right| \right)
\label{kin}
\end{equation}
where 
\begin{equation}
 \N_{inj} (\gamma) = \frac{1}{t} \frac{dN_l}{d\gamma} 
\end{equation}
is the rate at which leptons are created, calculated numerically from equation (\ref{num2}) that gives 
the distribution of leptons formed over a dynamical timescale $t$, and
\begin{displaymath}
 \frac{d\gamma}{dt} = \frac{dt'}{dt} \frac{d\gamma}{dt'} = -\frac{2\Gamma}{z+1} \frac{4 \sigma_e}{3m_e c}  
    (\gamma^2 -1) u'_B (Y+1) 
\end{displaymath}
\begin{equation}
  = - k_r (\gamma^2 -1) \;,\; k_r \equiv \frac{1}{3 \pi (z+1)} \frac{\sigma_e}{m_e c} \Gamma B^2 (Y+1)
\label{rad}
\end{equation}
is the radiative cooling rate of pairs, $t'$ being the shock-frame time, $u'_B = B^2/8\pi$ the
shock-frame magnetic energy density, and
\begin{equation}
  Y = \frac{4}{3} \int_1^\infty \gamma^2 d\tau_{sc} = \frac{\sigma_e}{3 \pi R^2} 
   \int_1^\infty \gamma^2 \N(\gamma) d\gamma
\end{equation}
is the Compton parameter (the ratio of the inverse-Compton to synchrotron losses), $\tau_{sc} =
\sigma_e N/(4\pi R^2)$ being the optical thickness to photon scattering by $N$ leptons in a source
of radius $R$.

 The lepton distribution $\N_{inj}$ depends on the properties of the high-energy emission
($\Phi, \veps_b, \alpha, \beta$) and of the photon source ($\Gamma, R$). Figure 3 shows the instantaneous
and integrated injected lepton distributions $\N_{inj}$ for a source with constant $\Phi, \alpha, \beta$,
a source deceleration corresponding to a blast-wave interacting with a wind-like medium -- 
$\Gamma = \Gamma_o (t/t_o)^{-1/4}$, and an evolution of the high-energy spectrum break as expected for
the forward-shock emission: $\veps_b = \veps_b(t_o) (t/t_o)^{-3/2}$. Because $\veps_b < \tveps_b$ 
(equation \ref{taulow}), the photon front is optically-thick to pair-formation above $\veps_-$ 
(equation \ref{epm}), thus the pairs have a $\N_{inj} \propto \gamma^{-\beta}$ distribution 
above $\gamma(\eps'_-)$ (equation \ref{dN2}), which increases with time: $\gamma(\eps'_-) \propto
\Gamma^5 t^2 \propto t^{3/4}$.

\begin{figure*}
\centerline{\psfig{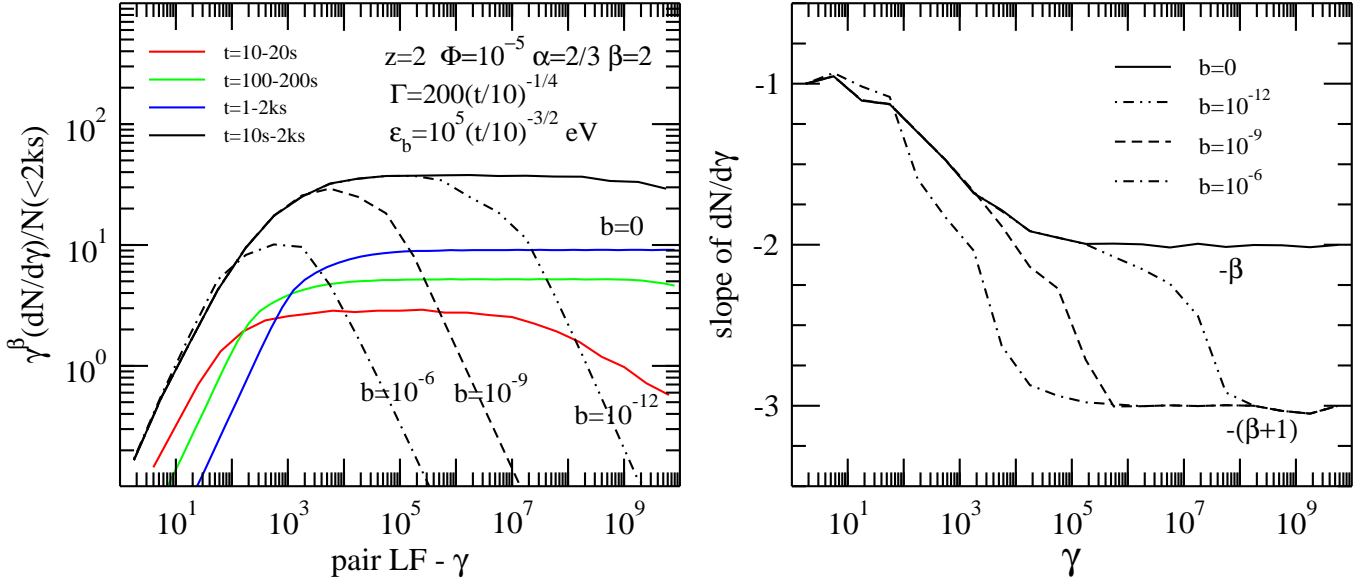}} 
\figcaption{{\sl Solid lines} show the distribution of leptons formed at three epochs (10 s, 100 s, 1 ks - 
   colored lines) and total distribution at 2 ks (black), integrated over the indicated source evolution
   and without radiative cooling ($b=0$). The quasi-instantaneous distributions are normalized to that 
   integrated up to 2 ks. The source GeV fluence is constant, its deceleration $\Gamma (t)$ and break 
   energy $\veps_b$ evolutions correspond to a blast-wave interacting with a wind-like external medium. 
   {\sl Black lines} show time-integrated (up to 2 ks) lepton distribution with synchrotron and inverse-Compton 
   cooling, for different fractions $b$ of the post-shock energy that is in the magnetic field.  
   The cooled lepton distributions display a break at the energy for which leptons lose half of their energy 
   over a dynamical timescale. Above the break, the lepton distribution slope increases by unity (right panel).
   The higher the magnetic field parameter $b$, the lower the lepton cooling-break energy.  }
%Fig3
\end{figure*}

 Figure 3 also shows the lepton distribution resulting from pair-formation with the above properties
and cooled radiatively by a magnetic field 
\begin{equation}
  \frac{B^2}{8\pi} = 4\, (b n) \, \Gamma^2 m_p c^2  
\label{Bn}
\end{equation}
that is a fraction $b$ of the post-shock energy.
The cooled lepton distribution develops a break at an energy $\gamma_c$ that 
decreases with time, with the lepton distribution being that injected at $\gamma < \gamma_c$ and
having a slope larger by 1 than that injected for $\gamma > \gamma_c$. 

 That feature for the radiative cooling of a power-law distribution of particles can be derived from
the kinetic equation for particle cooling (equation \ref{kin}), rewritten as
\begin{equation}
  \frac{\partial \N}{\partial t} = k_i \gamma^{-\beta}  + k_r \frac{\partial (\gamma^2 \N)}{\partial \gamma} 
\end{equation}
using equation (\ref{rad}). Trying a power-law solution $\N = a(t) \gamma^{-p}$, leads to
\begin{equation}
 k_i \gamma^{-\beta} = \frac{da}{dt} \gamma^{-p} + (p-2) k_r a \gamma^{1-p}
\end{equation}

 If the first term on the right-hand side is dominant (i.e. $da/dt \gg k_r a \gamma$ and the 
radiative cooling term is negligible), then $p = \beta$ and $da/dt = k_i$, thus $a = \int k_i dt$ 
and $\N = (\int k_i dt) \gamma^{-\beta}$, hence the effective distribution is the 
integrated injected distribution, at energies $\gamma$ that satisfy $\gamma \ll \gamma_c^{(1)}$, where
\begin{equation}
 \gamma_c^{(1)} \equiv \frac{da/dt}{k_r a} = \frac{k_i}{k_r \int k_i dt} 
\end{equation}
If the second, radiative cooling term on the rhs is dominant, then $p = \beta + 1$ and $a \simeq k_i/k_r$,
hence $\N = (k_i/k_r) \gamma^{-(\beta+1)}$. This solution exists for $|da/dt| \gg k_r a \gamma$, 
which is equivalent to $\gamma \gg \gamma_c^{(2)}$, where 
\begin{equation}
 \gamma_c^{(2)} \equiv \frac{|da/dt|}{k_r a} = \frac{1}{k_i} \left| \frac{d}{dt} \frac{k_i}{k_r} \right|
\end{equation}
The above two cooling energies are comparable
\begin{equation}
 \frac{\gamma_c^{(1)}}{\gamma_c^{(2)}} = \frac{k_i t}{\int k_i dt} \left/ \frac {d\ln |k_i/k_r|} {d\ln t} \right.
\end{equation}
if $k_i$ does not vary too fast and if $k_i/k_r$ is moderately evolving in time.
Thus, the effective lepton distribution is
\begin{equation}
 \N (\gamma) = \left\{ \begin{array}{ll}  (\int k_i dt) \gamma^{-\beta} & \gamma \ll  \gamma_c^{(1)} \\
        \frac{\displaystyle k_i}{\displaystyle k_r} \gamma^{-(\beta+1)} & \gamma_c^{(2)} \ll \gamma
    \end{array}\right.
\label{Np}
\end{equation}
For a constant injection rate $k_i$, the $\gamma_c^{(1)}$ becomes simpler
\begin{equation}
 \gamma_c \equiv \frac{1}{k_r t} \propto \frac{1}{\Gamma B^2 (Y+1) t} 
\end{equation}
and the effective distribution can be approximated as
\begin{equation}
 \N (\gamma)  \simeq \N (\gamma_c) \left\{ \begin{array}{ll}  
        \left(\frac{\displaystyle \gamma_c}{\displaystyle \gamma}\right)^\beta & \gamma < \gamma_c \\
        \left(\frac{\displaystyle \gamma_c}{\displaystyle \gamma}\right)^{\beta+1} & \gamma_c < \gamma 
    \end{array}\right. \N (\gamma_c) \equiv \frac{k_i}{k_r} \gamma_c^{-(\beta+1)}
\end{equation}

 We note that the cooled lepton distribution is not calculated from equation (\ref{kin}) because it
is unstable and "suffers" from a Courant-like condition, with the timestep $\delta t$ being upper-limited
by the cooling time $t_c (\gamma) = (k_r \gamma)^{-1}$ of the highest energy leptons $\gamma_{max}$ in the 
calculation, which can be $t_c \ll t$ for $bn \gamma_{max} > 10^3 \Gamma_{2.3}^{-3} (Y+1)^{-1}$. 
Instead, the lepton distribution is calculated numerically by tracking the flow of particles on a 
1-dimensional energy grid, using the cooling law of equation (\ref{rad}) 
\begin{equation}
 \frac{\gamma + 1}{\gamma - 1} = \frac{\gamma_o + 1}{\gamma_o - 1} e^{2 k_r \delta t} \;,\;
 \frac{1}{\gamma} \stackrel{\gamma \gg 1}{=} \frac{1}{\gamma_o} + k_r \delta t
\end{equation}
with $\gamma$ the energy of a lepton that had initially an energy $\gamma_o$, after a timestep $\delta t$.
Pair-energy tracking means accounting for that \\ 
(1) a fraction $\min [1, \delta t/t_c(\gamma)]$ of the $\delta N = \N(\gamma) \delta \gamma$ leptons
  existing in a cell $(\gamma,\gamma + \delta \gamma)$ exit that cell due to their cooling during $\delta t$,  \\
(2) a fraction $\max [1, t_c(\gamma)/\delta t]$ of the leptons $\delta N_{inj} = \N_{inj} (\gamma) 
\delta \gamma \delta t$ injected in a cell remain in that cell after cooling for $\delta t$, \\
(3) leptons existing in all cells between energies $\gamma_o (\gamma)$ and $\gamma_o (\gamma +\delta \gamma)$
cool during $\delta t$ to cell $(\gamma,\gamma + \delta \gamma)$, \\
(4) a fraction $[t_c(\gamma_o \rightarrow \gamma) - t_c(\gamma_o \rightarrow \gamma +\delta \gamma)]/\delta t$
of the leptons injected at energy $\gamma_o$ cool to cell $(\gamma,\gamma + \delta \gamma)$ during $\delta t$,
with $t_c(\gamma_o \rightarrow \gamma) \equiv (\gamma^{-1} - \gamma_o^{-1})/k_r$, the cooling time from 
energy $\gamma_o$ to $\gamma$. \\

\section{Radiation Emission}

 The calculation of the synchrotron self-Compton from pairs three components: synchrotron emission,
synchrotron self-absorption, and inverse-Compton scattering of the self-absorbed synchrotron spectrum.
We consider only the first inverse-Compton scattering, which is appropriate approximation when the Compton 
parameter $Y$ is sub-unity and a necessary approximation when leptons radiating at the observing frequency 
$\nu$ cool faster than they are created. The former case requires a magnetic field in the shock fluid that 
is not much below equipartition ($b \simg 10^{-3}$); for the latter case, the argument is that, in an source 
that is optically-thin to electron scattering, the lepton distribution can change substantially during the time
it takes a photon to cross the source and be scattered, hence the lepton distribution that produced the
seed photon is not the same as the lepton distribution that upscatters it, an a time-dependent treatment
of upscatterings is needed. For $Y > 1$, ignoring higher-order scatterings leads to an underestimation of 
the flux at higher energies (above X-rays) and an overestimation of the synchrotron and first inverse-Compton 
flux above the cooling frequency (which would be lower if higher-order scatterings were accounted).

\subsection{Synchrotron Emission}

 For a relativistic source moving at Lorentz factor $\Gamma$, the received synchrotron flux from a distribution
of leptons $\N (\gamma)$ at observer frequency $\nu$ is
\begin{equation}
 F_{sy} (\nu) = \frac{z+1}{4 \pi d_l^2} \Gamma \int d\gamma \N (\gamma) P'_{sy} 
      \left( \frac{\displaystyle z+1}{\displaystyle \Gamma} \nu, \gamma \right) 
\label{Fsy1}
\end{equation}
where the relativistic boost of the comoving-frame emission at frequency $\nu' = (z+1)\nu/\Gamma$ gets
only one power of $\Gamma$ from the contraction of photon arrival time $dt = dt_{lab}/\Gamma^2 = dt'/\Gamma$
relative to the comoving-frame emission time $dt'$ (the boost $\Gamma$ in photon energy is "lost" to that
a comoving energy range $d\nu'$ is stretched into $\Gamma d\nu'$ for the observer, and the angular beaming
boost $\Gamma^2$ is "lost" because, for the observer, that beaming reduces the solid angle of a spherical 
source by a factor $\Gamma^2$), and 
\begin{equation}
 P'_{sy} (\nu', \gamma) = \frac{e^3 B}{m_ec^2} f_{sy} \left( \frac{\nu'}{\nu'_{sy} (\gamma)}  \right)
\label{psy}
\end{equation}
is the specific synchrotron power for a lepton, with $f_{sy}$ the "synchrotron function" and
\begin{equation}
 \nu'_{sy}(\gamma) = \frac{3}{16} \frac{e}{m_e c} B \gamma^2 = 3.3 \times 10^6 B_o \gamma^2 \; {\rm Hz}
\label{nusy}
\end{equation}
is the synchrotron characteristic frequency at which a lepton of energy $\gamma$ radiates most of its
emission. The synchrotron function is an integral over the modified Bessel function of 5/3 order and has
the following asymptotic behavior
\begin{equation}
 f_{sy} (x) \simeq \left\{ \begin{array}{ll}  1.71\, x^{1/3} & x \ll 1  \\
     1.25\, x^{1/2} e^{-x} & x \gg 1 \end{array} \right.
\end{equation}
This approximation would be useful if the synchrotron emission at frequency $\nu'$ were produced by
leptons whose characteristic synchrotron frequency $\nu'_{sy}$ is far from $\nu'$, however, the opposite
is true. We approximate the synchrotron function with the asymptotic behaviors given above but with 
coefficients such that 
$i)$ $f_{sy}$ is continuous at $x=1/2$ (where $x^{1/2} e^{-x}$ has a maximum) and
$ii)$ its integral is equal to that of the exact synchrotron function, $\int f_{sy}(x) dx = (4/3)^3$,
yielding a power-per-lepton $P'(\gamma) = \int P'_{sy}(\nu',\gamma) d\nu' = (4/3) \sigma_e c \gamma^2 
(B^2/8\pi)$. The following approximation
\begin{equation}
 f_{sy} (x) \simeq \left\{ \begin{array}{ll}  1.50\, x^{1/3} & x < 1/2  \\
     2.77\, x^{1/2} e^{-x} & x > 1/2 \end{array} \right.
\label{fsy}
\end{equation}
satisfies the above constraints and has a maximum $f_{sy} (0.5) = 1.2$ (that of the exact synchrotron 
function is $f_{sy} (0.3) = 0.92$).

 Substituting equations (\ref{psy}) and (\ref{nusy}) in (\ref{Fsy1}), we get
\begin{equation}
 F_{sy} (\nu) = \frac{4.3 \times 10^{-56}}{(z+1)^3} B \Gamma \int d\gamma \N(\gamma) 
       f_{sy} \left( \frac{\gamma_\nu^2}{\gamma^2} \right)  \; {\rm Jy}
\label{Fsy2}
\end{equation}
in cgs units, where
\begin{equation}
 \gamma_\nu \equiv 5.5 \times 10^{-4} \left[ \frac{(z+1)\nu}{\Gamma B} \right]^{1/2}
\end{equation}
satisfies $\nu_{sy} (\gamma_\nu) = \nu$. Owing to the exponential cut-off of the synchrotron function
at $x > 1$, only leptons with energy above $\gamma_\nu$ produce the synchrotron emission at frequency $\nu$. 
Then, approximating the synchrotron function by only its $x < 1/2$ branch, one obtains 
\begin{equation}
 F_{sy} (\nu) = \frac{0.92 \times 10^{-57}}{(z+1)^{8/3}} (B^2 \Gamma^2 \nu)^{1/3}
      \int_{\sqrt{2}\gamma_\nu}^\infty d\gamma \frac{\N(\gamma)}{\gamma^{2/3}}  \; {\rm Jy}
\end{equation}
For a power-law distribution of particles, $\N(\gamma) \propto \gamma^{-p}$, this leads to the
well-known spectrum $F_{sy}(\nu) \propto \nu^{-(p-1)/2}$.

\subsection{Synchrotron Self-Absorption}

 Starting from equation (6.49) of Rybicki \& Lightman (1949), taking into account that the lepton
column-density is $N/4\pi R^2$, and using the synchrotron emissivity per lepton given in equation (\ref{psy}), 
the synchrotron self-absorption optical thickness at observer frequency $\nu = \Gamma \nu'/(z+1)$ is 
\begin{equation}
  \tau_a (\nu) = - \frac{1}{32 \pi^2 R^2} \left[ \frac{\Gamma}{(z+1) \nu} \right]^2 \frac{e^3 B}{(m_e c)^2} 
\end{equation}
\begin{displaymath}
     \times   \int d\gamma \gamma^2 f_{sy} \left( \frac{\gamma_\nu^2}{\gamma^2} \right)
       \frac{\partial}{\partial \gamma} \left( \frac{\N (\gamma)}{\gamma^2}  \right)
\end{displaymath}
Quick progress toward a simpler form can be made if we retain only the $ x < 1/2$ branch of the synchrotron 
function (equation \ref{fsy}) and set $f_{sy}(x > 1/2) = 0$ (motivated by the exponential cut-off):
\begin{equation}
  \tau_a (\nu) \simg - \frac{0.0083}{R^2} \left[ \frac{\Gamma}{(z+1) \nu m_e c} \right]^{5/3} (e^4 B)^{2/3} 
\end{equation}
\begin{displaymath}
  \times \int_{\sqrt{2}\gamma_\nu}^\infty d\gamma \gamma^{4/3} \frac{\partial}{\partial \gamma} 
   \left( \frac{\N (\gamma)}{\gamma^2} \right)
\end{displaymath}
The integral above can be re-written as
\begin{equation}
  \tau_a (\nu) \simg \frac{4.7}{R^2} \left[ \frac{\Gamma}{(z+1) \nu} \right]^{5/3} B^{2/3}
\label{tauabs}
\end{equation}
\begin{displaymath}
  \times \left( \left. \frac{\N}{\gamma^{2/3}} \right|_{\sqrt{2}\gamma_\nu} + 
  \frac{4}{3} \int_{\sqrt{2}\gamma_\nu}^\infty d\gamma \frac{\N}{\gamma^{5/3}} \right)
\end{displaymath}
in cgs units.
For a power-law distribution of leptons with energy, $\N (\gamma) \propto \gamma^{-p}$, the ratio
of the two terms above is $(3p+2)/4$, where $p = \beta,\beta+1$ (equation \ref{Np} and Figure 3).
For a typical spectrum of the high-energy photons, $\beta \simeq 2$, this ratio is 2 or 11/4,
thus the first term is dominant. Either term is proportional to $\gamma_\nu^{-(p+2/3)}$ and
yields $\tau_a (\nu) \propto \nu^{-(p+4)/2}$.

 The emerging synchrotron flux is the intrinsic flux (given in equation \ref{Fsy2}) reduced by the
escaping fraction (equation \ref{g}):
\begin{equation}
  F_{sy}^{(obs)} (\nu) =  \frac{1-e^{-\tau_a(\nu)}}{\tau_a(\nu)} F_{sy} (\nu)
\label{Fssa}
\end{equation}
Below the self-absorption frequency $\nu_a$, where $\tau_a(\nu_a) = 1$, the received synchrotron flux
satisfies $F_{sy}^{(obs)} (\nu) = F_{sy}(\nu)/\tau_a(\nu)$ which, for a power-law distribution
of particles, is $F_{sy}^{(obs)} (\nu < \nu_a) \propto \nu^{-(p-1)/2}/\nu^{-(p+4)/2} = \nu^{5/2}$, 
another well-known result.

\subsection{Inverse-Compton Emission}

 For a lepton of energy $\gamma$ scattering a photon of frequency $\nu'_o$ in the Thomson regime, 
the inverse-Compton emissivity at photon frequency $\nu'$ is (equation 7.26a in Rybicki \& Lightman 1979)
\begin{equation}
  j'_{ic} (\nu') =  \frac{3 \sigma_e F'_o}{4 (\gamma \nu'_o)^2} 
      f_{ic} \left( \frac{\nu'}{4\gamma^2 \nu'_o} \right) \nu'
\end{equation}
where $F'_o$ is the comoving-frame energy flux of $\nu'_o$ photons and 
\begin{equation}
  f_{ic} (x) = \left\{ \begin{array}{ll} 2x \ln x -2 x^2 + x +1 &  (16\gamma^4)^{-1} < x < 1 \\ 
    0 & {\rm otherwise} \\ \end{array} \right.
\end{equation}
 By integrating $\N (\gamma) (\nu'_o) j'_{ic} (\nu')$ over the lepton distribution and over the 
incident synchrotron photon spectrum, one obtains the comoving-frame inverse-Compton luminosity, 
which is enhanced by a factor $\Gamma$ to yield the lab-frame inverse-Compton luminosity.
Then, the received inverse-Compton emission from pairs at frequency $\nu = \Gamma \nu'/(z+1)$ is
\begin{equation}
  F_{ic} (\nu)  = \frac{3(z+1)}{8 \pi d_l^2} \sigma_e \Gamma \nu' \int d\gamma \N(\gamma) \int d\nu'_o 
    \frac{F'_o (\nu'_o)}{(\gamma \nu'_o)^2} f_{ic} \left( \frac{\nu'}{4\gamma^2 \nu'_o} \right) 
\label{mccoy}
\end{equation}
The comoving-frame flux of incident photons $F'_o (\nu'_o)$ in a source of radius $R$ is related to the 
received synchrotron flux at photon frequency $\nu_o = \Gamma \nu'_o/(z+1)$ through
\begin{equation}
 F_{sy} (\nu_o) = \frac{z+1}{4 \pi d_l^2} \Gamma (4 \pi R^2) F'_o (\nu'_o)
\end{equation}
Substituting in equation (\ref{mccoy}), and changing from comoving-frame to observer-frame photon frequencies,
it follows that the received inverse-Compton flux can be calculated from the received synchrotron flux 
(equation \ref{Fsy2}):
\begin{equation}
  F_{ic} (\nu)  = \frac{3\sigma_e}{16 \pi R^2} \nu \int \frac{d\gamma}{\gamma^2} \N(\gamma)
      \int \frac{d\nu_o}{\nu_o^2} F_{sy}^{(obs)} (\nu_o) f_{ic} \left( \frac{\nu}{4\gamma^2 \nu_o} \right) 
\label{Fic1}
\end{equation}
where $F_{sy}^{(obs)}$ is the self-absorbed synchrotron flux (equation \ref{Fssa}), because synchrotron 
self-absorption reduces the flux of photons incident on a scattering lepton in the same way that it
affects the received synchrotron flux.
 
 An approximation that changes very little the spectrum of the inverse-Compton pair emission is that
where photons of frequency $\nu$ result from the upscattering by a lepton of energy $\gamma$ of {\sl only}
synchrotron photons of frequency $\nu_o = 3 \nu/(4 \gamma^2)$, which is motivated by that the average
energy of an upscattered photon is $(4/3) \gamma^2 \nu_o$. This is equivalent to approximating the
inverse-Compton function with a $\delta$-function
\begin{equation}
 f_{ic} \left( \frac{\nu}{4\gamma^2 \nu_o} \right) = \delta \left( \frac{3\nu}{4\gamma^2 \nu_o} - 1 \right)
\end{equation}
Then, equation (\ref{Fic1}) becomes
\begin{equation}
  F_{ic} (\nu)  = \frac{\sigma_e}{4\pi R^2} \int d\gamma \N(\gamma) \left[ F_{sy}^{(obs)} 
                    \left(\frac{3\nu}{4\gamma^2} \right) \right] 
\label{Fic2}
\end{equation}
\begin{displaymath}
                = \int d\tau_e \left[ F_{sy}^{(obs)} \left( \frac{3\nu}{4\gamma^2} \right) \right]
\end{displaymath}
where $\tau_e = (\sigma_e N_l)/(4\pi R^2)$ is the pairs' optical-thickness to photon scattering.
Equation (\ref{Fic2}) means that the inverse-Compton specific flux is the integral over the lepton
distribution of the scattered synchrotron flux.

\vspace*{6mm}
\section{Emission from pairs formed in the shocked fluid}

\subsection{Approximate dependences}

\begin{figure*}
\centerline{\psfig{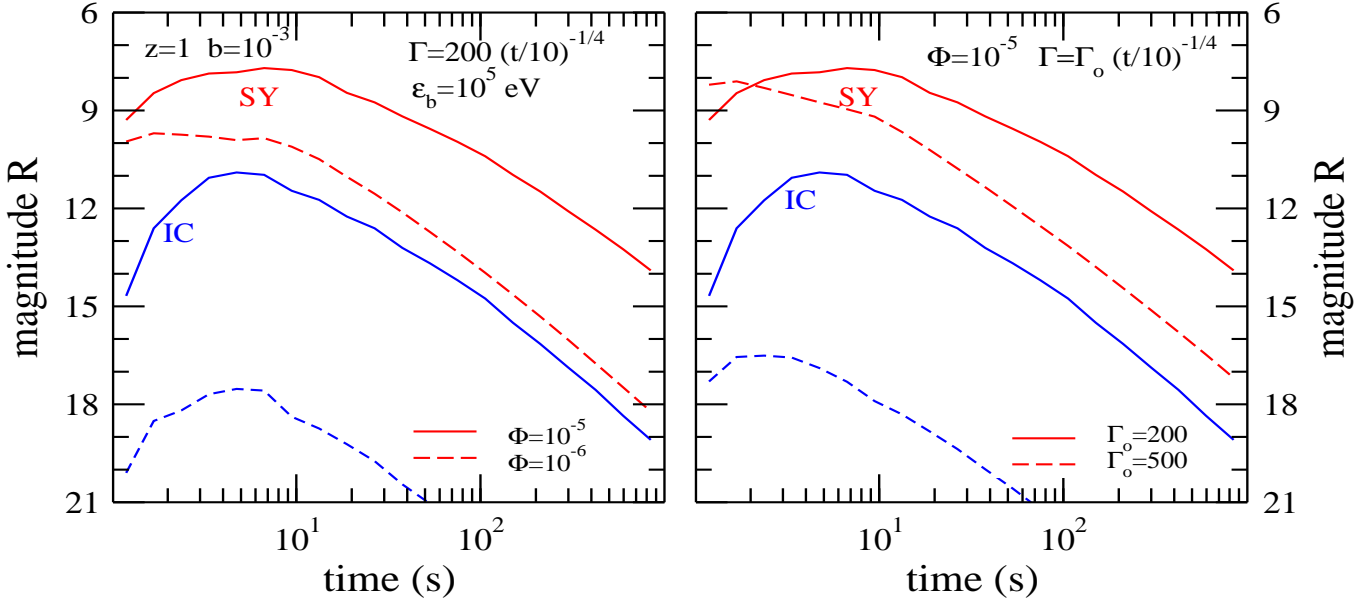}}
\figcaption{ Dependence of pair emission on the fluence of the high-energy LAT afterglow fluence $\Phi$
   (left panel) and on the GeV initial Lorentz factor $\Gamma_o$ (right panel), with other parameters
   as indicated. Red lines are for synchrotron and blue lines for inverse-Compton.  }
%fig4
\end{figure*} 

 The most important parameters that determine the pair emission are those that set the number of pairs 
-- the blast-wave initial Lorentz factor $\Gamma_o$ and the afterglow high-energy fluence $\Phi$ --
and the magnetic field -- the $nb$ product (equation \ref{Bn}). Less effective, but still relevant,
are three other parameters that determine the number of pairs: the slopes $\alpha$ and $\beta$ of the 
high-energy spectrum, and its break energy $\veps_b$. 

 Equations (\ref{Fsy2}) and (\ref{Fic2}) suggest the following dependences for the synchrotron and 
inverse-Compton flux from pairs:
\begin{equation}
 F_{sy} (\nu) \propto B\Gamma N_\pm \;,\; F_{ic} (\nu) \propto \frac{N_\pm}{R^2} F_{sy} 
         \propto \frac{N_\pm^2}{R^2} B \Gamma
\end{equation}
where $R \propto \Gamma^2 t$ (equation \ref{R}), $B \propto \sqrt{nb} \Gamma$ (equation \ref{Bn}),
and $N_\pm \propto \Phi^2/(\Gamma^6 t^2)$ (equations \ref{N1} and \ref{N2}). 
Equation (\ref{Fsy2}) actually means that $F_{sy} (\nu) \propto N(>\gamma_\nu)$, hence the use of
$N_\pm$ here is accurate only when the number of pairs above $\gamma_\nu$ is a fixed fraction of the 
total number of pairs. That is satisfied only above the pair cooling-energy, where pairs produced 
during one cooling timescale reside. With the above substitutions, we find that
\begin{equation}
 F_{sy} (\nu) \propto \frac{\Phi^2 (nb)^{1/2}}{\Gamma^4 t^2} \;,\; 
 F_{ic} (\nu) \propto \frac{\Phi^4 (nb)^{1/2}}{\Gamma^{14} t^6}
\end{equation}
For a wind-like medium, where $n \propto R^{-2}$ and $\Gamma \propto t^{-1/4}$, we arrive at
\begin{equation}
 F_{sy} (\nu) \propto \frac{\Phi^2 \sqrt{b}}{\Gamma^6 t^3} \propto \Phi^2 t^{-3/2}  \;,\; 
 F_{ic} (\nu) \propto \frac{\Phi^4 \sqrt{b}}{\Gamma^{16} t^7} \propto \Phi^4 t^{-3}
\label{decay}
\end{equation}

 This shows that the inverse-Compton flux has a very strong dependence on the high-energy afterglow fluence 
and a super-strong dependence on the Lorentz factor of the GeV afterglow, which suggest that inverse-Compton 
emission from pairs could be relevant (i.e could overshine the synchrotron flux) only for the brightest 
LAT afterglows ($\Phi \sim 10^{-5} \ergcm2$) and for the slowest GeV sources in which pairs are still 
optically-thin ($\Gamma \simg 130$ - equation \ref{Gtau}). 

% Given that the cooled pair distribution with energy has slopes $\beta$ and $\beta+1$ (equation \ref{Np} 
%and Figure 3) below and above the cooling pair energy, with $\beta$ the slope of the LAT afterglow spectrum, 
%it follows that the spectrum of the pair emissions has slopes $F_{sy/ic} (\nu) \propto \nu^{-(\beta-1)/2}$
%below the cooling frequency and steeper by 1/2 above it. For $\beta = 2$, the pair emission spectrum
%is $F_{sy/ic}(\nu) \propto \nu^{-1/2}, \nu^{-1}$. Noting that number of pairs has a weak dependence
%on the slope $\beta$, it follows that the brightness of the pair emission should not be correlated with
%the slope of the LAT spectrum (another {\sl observable}). The same is true for the decay rate of pair light-curves.

\subsection{Light-curves and spectra}

 The pair light-curves shown in Figure 4 illustrate the correlation of the pair flux with the observable 
LAT fluence $\Phi$ and the unknown source Lorentz factor $\Gamma_o$ at the peak epoch $t_o$ of the LAT 
light-curve. Those light-curves were obtained by integrating the synchrotron and inverse-Compton fluxes 
given in equations 
(\ref{Fsy2}), (\ref{tauabs}), (\ref{Fssa}), and (\ref{Fic1}) (or \ref{Fic2}), over the deceleration of a 
blast-wave interacting with a massive-star wind. Although $N_\pm \propto \Phi^2/(\Gamma_o^6 t_o^3)$ is
satisfied by the numerical calculation of the pairs formed, the numerical pair fluxes display a weaker 
correlation with $\Phi$ and $\Gamma_o$ (and also with $t_o$) than given in equation (\ref{decay}), which 
is due to the use of $N_\pm$ in the derivation of that equation.  

 Equations (\ref{N1}) and (\ref{N2}) show that the number of pairs is weakly dependent on the
unknown break energy $\veps_b$ of the LAT spectrum (another model {\sl parameter}), with more pairs being 
formed for a lower $\veps_b$, because that increases the optical thickness to pair formation (equations 
\ref{tau2} and \ref{epm}).
The brightness of the LAT high-energy spectral component at sub-MeV photon energies relative to that
of the burst is the criterion for choosing the two prescriptions given in Figure 5 for the unknown 
break-energy: GRBs with a fast-decaying tail require that the $\veps_b$ of a bright LAT component
remains above MeV for the duration of the tail, while bursts with a slowly-decaying tail allow lower
$\veps_b$ (decreasing or not).
As expected, a lower $\veps_b$ yields a brighter pair emission, and a decreasing $\veps_b$ leads to a
slower decay of the pair light-curve. The latter behavior provides a criterion for identifying early 
optical afterglows produced by pairs: slowly-dimming pair afterglows (due to a decreasing $\veps_b$) 
cannot follow fast-falling bursts (which are incompatible with a decreasing $\veps_b$). However, 
fast-falling pair afterglows can follow either type of burst tail (fast or slowly decreasing).

\begin{figure*}
\centerline{\psfig{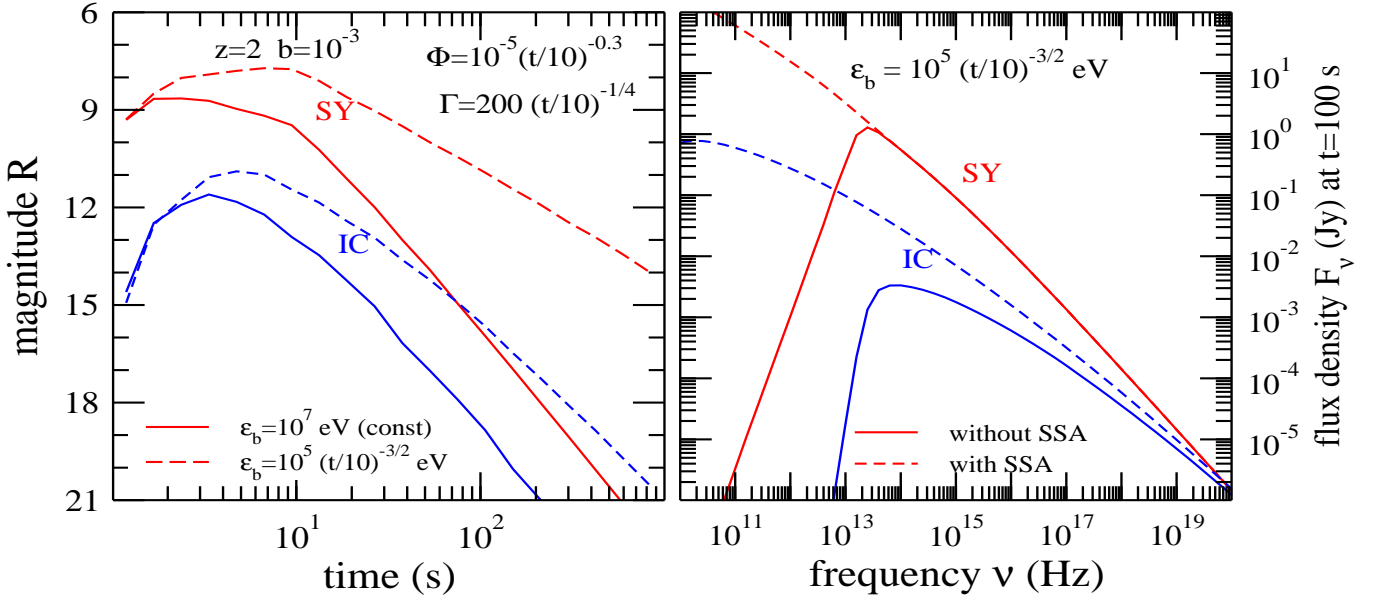}}
\figcaption{ Left panel: optical light-curves from pairs formed in the shocked fluid, for the indicated
    parameters, and for the wind-like external medium given in equation (\ref{WR}). Solid lines are
    for a constant break-energy $\veps_b$ of the high-energy emission that forms pairs. The synchrotron
    flux (red line) exhibits a decay that is slightly steeper than estimated from equation (\ref{decay}), 
    while the inverse-Compton flux (blue line) has a decay that is significantly slower. Dashed lines 
    are for a decreasing $\veps_b$, which yields a slower flux decay.
    Right panel: synchrotron and inverse-Compton spectra at $t = 100$ s, for a decreasing $\veps_b$.
    Dotted lines show spectra without accounting for synchrotron self-absorption (SSA), solid lines are for
    absorbed spectra.
    Owing to the large magnetic field parameter $b$, the Compton parameter is below unity and the 
    inverse-Compton emission is dimmer than synchrotron at all frequencies of interest. The synchrotron
    cooling break is slightly below the optical.  }
%fig5
\end{figure*}

\subsection{Application to GRB 130427A}

 Figure 6 shows a fit to the super-bright optical flash of GRB 130427A (RQD2/Raptor - Vestrand et al 2014) 
with the synchrotron emission from internal pairs formed from the high-energy emission monitored by LAT
(Fan et al 2013, Tam et al 2013, Ackermann et al 2014). Observations set the high-energy fluence $\Phi$ 
and the spectral slope $\beta$ above the unknown break energy $\veps_b$, which is a free model parameter. 
The initial Lorentz factor $\Gamma_o$ of the high-energy photons source and the magnetic field parameter 
$b$ in the shocked fluid are two other model parameters. 

 As indicated by equation (\ref{decay}), with the high-energy fluence $\Phi$ set by observations,
the brightness of the optical flash of 130427A constrains the combination $b/\Gamma_o^{12}$. 
We find that $\Gamma_o < 300$ is required to match the brightness of the 130427A optical flash because,
for higher Lorentz factors, the number of pairs formed is too small, and the maximal optical flux from
pairs, obtained for a magnetic field that brings the peak of the self-absorbed synchrotron spectrum
in the optical, falls short of the peak brightness of GRB 130427A's optical counterpart.

 The decay of the pair synchrotron optical light-curve depends primarily on the LAT light-curve $\Phi(t)$
and on the blast-wave deceleration $\Gamma(t)$. As those quantities are already set, the optical flash decay 
($F_o \propto t^{-2}$) constrains the slope of the LAT spectral component below its $\veps_b$ break and 
the evolution of $\veps_b$. 
 
 For $\alpha=2/3$ (i.e. $\veps_b$ is the peak energy of a synchrotron or inverse-Compton spectrum without cooling), 
we find that $\veps_b \propto t^{1/2}$ is required to match the optical light-curve decay at 10--100 s. 
This time-dependence is consistent with the behavior of the cooling-break of the synchrotron spectrum 
from the forward-shock (and for a wind-like medium), but $\alpha=2/3$ is inconsistent with the expected 
value $\alpha = \beta - 1/2$ in that case. 

 For $\alpha = 3/2$ (when $\veps_b$ would be the injection peak of the sy/ic spectrum with electron cooling) 
or for $\alpha = \beta -1/2 = 1.7$ (when $\veps_b$ would be the cooling-break of a sy/ic spectrum),
we find that the optical flash decay requires $\veps_b \propto t^2$, which is consistent with the evolution 
of the cooling-break of the inverse-Compton spectrum from the forward-shock (and for a wind-like medium). 
%For $\veps_b (t_o)= 500$ keV and $\alpha = \beta - 1/2 = 1.7$, the extrapolation of the LAT component to energies 
%below $\veps_b$ matches quite well the 10 keV Swift/BAT light-curve shown in Figure 6, but the prompt emission 
%measured for GRB 130427A is brighter than the extrapolation of the LAT component at 10 keV--1 MeV, and is
%also harder: $\alpha_{bat} = 1.21 \pm 0.02$ at 50--220 s and $\alpha_{konus} = 0.96 \pm 0.01$ at 0--19 s.

 Therefore, fits to the decay of the prompt optical emission of GRB 130427A with emission from internal-pairs
sets constraints on the unmeasured peak energy $\veps_b$ of the LAT spectral component that do not elucidate
its shock origin. Furthermore, numerical fits to the multiwavelength emission of this afterglow show comparable 
contributions to the LAT emission arising from both synchrotron reverse and forward shocks (Panaitescu et al 2013). 

 Figure 6 also shows that the X-ray emission from pairs and that LAT component contribution to the X-ray
are below the fluxes measured by Swift, and that the formation of enough pairs to produce a bright optical
flash does not entail a high attenuation of the LAT spectrum above 10 GeV. For the highest Lorentz factor 
$\Gamma_o$ that allows a good fit to the optical flash, the intrinsic power-law spectrum above 10 GeV is
attenuated by up to 50 percent at $t=10$ s (when attenuation is maximal), which is not inconsistent with 
the detection by LAT of a 70 GeV photon at 18 s. However, Lorentz factors $\Gamma _o \siml 200$ are incompatible
with that detection.

 Although a good fit with the internal-pair emission for the optical flash of GRB 130427A is obtained,
we do not propose this origin for the optical counterpart of GRB 130427A because modeling of the broadband
(radio, optical, X-ray, and GeV) emission of this afterglow (Panaitescu, Vestrand \& Wozniak 2013), 
from 100 s to tens of days, has shown that its wind-like ambient medium must be very tenuous, which leads 
to an initial Lorentz factor $\Gamma_o \simeq 750$ that is much higher than allowed  by fitting the optical 
flash with internal-pairs emission ($\Gamma_o \siml 300$). 

 Synchrotron emission from {\sl external} pairs formed ahead of the blast-wave from {\sl burst MeV} 
photons scattered by the ambient medium and, then, accelerated by the forward-shock can also produce 
a bright optical flash ($R < 10$) lasting for 100 s, provided that the initial source Lorentz factor 
is $\Gamma_o \sim 200$ (figures 4 and 7 of Kumar \& Panaitescu 2004). Vurm, Hascoet \& Beloborodov 
(2014) have found that the optical flash of GRB 130427A can be explained with synchrotron emission 
from external pairs accelerated by the forward-shock if that shock's Lorentz factor is a low $\Gamma = 200$.

 A similar model, but not investigated here, is the emission from the shock-accelerated {\sl external} pairs 
formed from {\sl afterglow MeV--TeV} photons ahead of the forward-shock. In one variant of that model --
pair-formation from {\sl unscattered} GeV photons -- the number of pairs is strongly decreasing with the 
source Lorentz factor, therefore it requires a low $\Gamma_o \siml 300$ to account for the optical flash of 
GRB 130427A. In the other variant -- pair-formation from GeV photons {\sl scattered} by the ambient medium
(which decollimates photons sufficiently to lower significantly the pair-formation threshold-energy and 
and enriches with pairs the medium ahead of the blast-wave) -- the number of pairs
should be less dependent on $\Gamma_o$. Owing to its similarity to the pair-wind formed from scattered
burst MeV photons, this model may also require a low $\Gamma_o$ to account for the optical flash of GRB 130427A.

 If all pair-based models for this flash require low Lorentz factors (for the seed-photon source)
that are incompatible with the afterglow $\Gamma$ inferred from multiwavelength data modeling, the
bright optical flash of GRB 130427A should be attributed to the reverse-shock (M\'esz\'aros \& Rees 1997)
that energizes some
incoming ejecta in an initial injection episode, followed by a quiet period when the ejecta electrons
cool radiatively and yield a fast-decaying flux, followed by a second, longer-lived injection episode,
during which the reverse-shock produces the optical emission measured for the early (up to few ks) 
afterglow of GRB 130427A (as proposed by Vestrand et al 2014).

\begin{figure*}
\centerline{\psfig{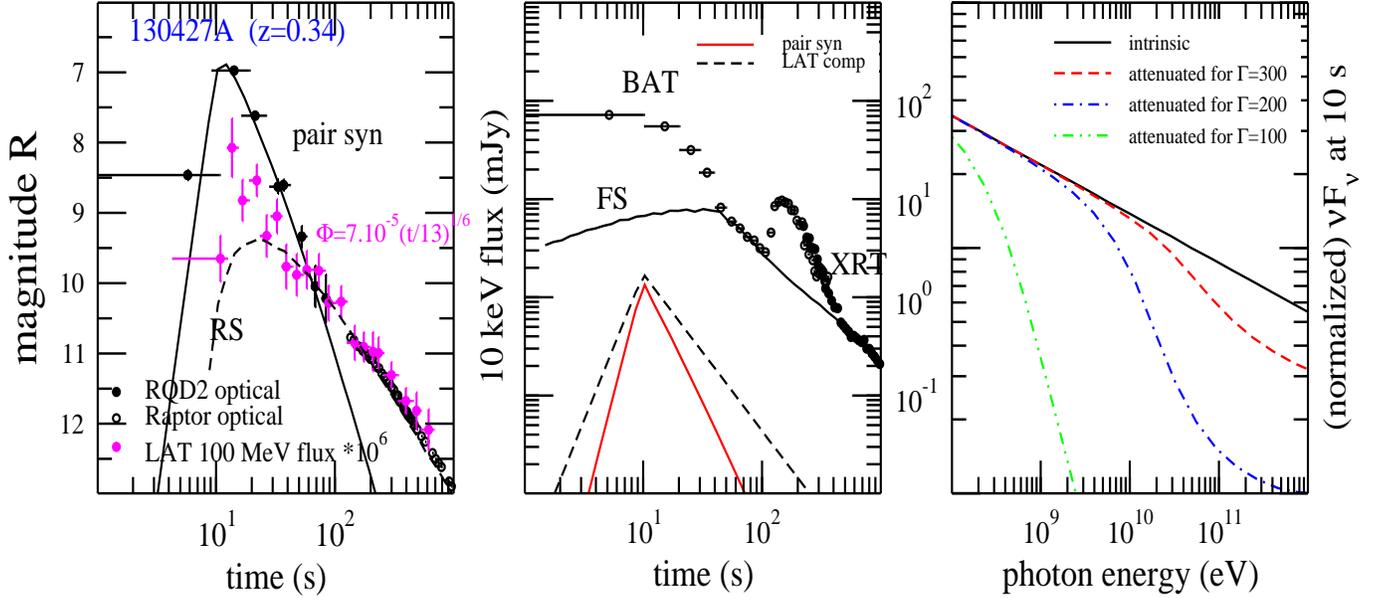}}
\figcaption{
   Left panel: early optical and GeV emission for GRB 130427A and a fit to the optical flash 
    (up to 100 s) with synchrotron emission from internal pairs formed in the shocked fluid. 
    LAT fluence (model input) after the $t_o = 10$ s peak is indicated; LAT spectral slope is 
    $\beta = 2.2$ (Tam et al 2013). 
    Model parameters are: source Lorentz factor before LAT peak $\Gamma_o = 300$, break-energy of the 
    LAT spectrum $\veps_b = 500\, (t/t_o)^{1/2}$ keV , magnetic field parameter $b = 10^{-3}$. 
    The burst ambient medium is that of equation (\ref{WR}), 
    which sets the source dynamics: $\Gamma (t) = \Gamma_o (t/t_o)^{-1/4}$. 
    After 100 s, the optical emission is well-fit by the reverse-shock emission (Panaitescu et al 2013).
   Mid panel: for the parameters of the optical flash fit, the pair emission is dimmer than the early 
    X-ray emission of this burst, monitored by Swift's BAT and XRT.
    The 10 keV emission from the LAT component with an assumed low-energy slope $\alpha = 2/3$ below 
    $\veps_b$ is also shown. Part of the burst tail and early X-ray afterglow can be explained with 
    the forward-shock emission. 
   Right panel: photon-photon attenuated spectra (dashed lines) at the LAT peak light-curve epoch,
    when attenuation is maximal, for the measured LAT spectrum (solid line) and various initial $\Gamma_o$.
    For $\Gamma_o = 300$, a moderate absorption occurs above 10 GeV, corresponding to a flux reduction 
    of at most 50 percent. For smaller $\Gamma_o$, attenuation is stronger.  
   }
%fig6
\end{figure*}

\section{Conclusions}

 In GRB afterglows, test photons of lab-frame energy above $\sim 10$ MeV form pairs in interactions
with target photons that are above the threshold for pair-formation. The number of pairs depends
moderately on the unknown break-energy $\veps_b$ of the high-energy component (LAT measures only photons 
above $\veps_b$), strongly on the afterglow GeV output (which is the observable LAT fluence $\Phi$), 
and very strongly on the Lorentz factor $\Gamma$ of the GeV source. 

 Below the radiative cooling break, the brightness of the synchrotron emission from (internal) pairs, formed 
in the shocked fluid (between the reverse and forward shocks), depends on their number (set by one observable 
-- $\Phi$ -- and two model parameters -- $\Gamma$ and $\veps_b$) and on the strength of the magnetic field 
between shocks (a third model parameter). For an intermediate/low $\Gamma$, pairs produce bright optical 
early afterglows even for a magnetic field that is several orders of magnitude below equipartition. In fact, 
strong magnetic fields do not warrant a much brighter optical emission because an enhanced radiative 
cooling reduces the number of pairs of sufficiently high energy to radiate synchrotron emission in the optical.
 
 The correlation between the number of pairs and the attenuation of the LAT spectrum, induced by the 
dependence of these two features on $\Gamma$, provides a way to identify optical counterparts that 
originate from internal pairs (formed in the GeV source). For the most relativistic afterglows 
($\Gamma \simg 500$), the internal-pairs emission should be dim and the LAT spectrum should be an 
unattenuated power-law, both because few pairs are formed. $300 \siml \Gamma \siml 500$ yields a moderately
bright optical flash and no detectable attenuation of the LAT spectrum. For the less relativistic afterglows 
($100 \siml \Gamma \siml 200$), when many pairs are formed, there should be a bright optical emission from 
pairs, accompanied by a significant attenuation of the LAT spectrum above 1 GeV. 

 An additional criterion for identifying optical counterparts from internal-pairs emission arises 
from that GRBs with fast-decaying tails require a peak energy $\veps_b \simg 10$ MeV of the LAT spectrum, 
which yields dimmer and faster-decaying optical emission from pairs. Slowly-decaying GRB tails do not 
exclude bright optical flash from pairs, hence there should be some correlation between the speed of the
GRB tail decay and the brightness of the pair optical flash.

%\acknowledgments{This work was supported by an award from the Laboratory Directed Research and
%   Development program at the Los Alamos National Laboratory} 

\end{document}